**Centralized adaptive traffic control strategy design across multiple intersections based on vehicle path flows: An approximated Lagrangian decomposition approach**

Peirong (Slade) Wang[a]

Email: pw685@msstate.edu

Pengfei (Taylor) Li[a] *

Tel: 662-325-7518; Email: pengfei.li@ msstate.edu

Farzana Chowdhury[a]

Email: frc36@msstate.edu

Li Zhang[a]

Email: lzhang@cee.msstate.edu

[a]Department of Civil and Environmental Engineering,

Mississippi State University, Mississippi State, MS, 39762, USA

* Corresponding author.

## Abstract

At the dawn of mobility as a service (MaaS), novel traffic data sources are viewed as a promising foundation in developing more adaptive and robust traffic control systems. It is expected that the vehicle dynamic origin-destination and path flows may become ubiquitously available in foreseeable future. The new type of traffic data will provide full-spectrum space-time characteristics of traffic dynamics among all intersections. Nonetheless, methodological gaps also exist toward that objective. To fully take the potential of new traffic data, novel traffic control representations will be necessary to model and control the space-time path flows across intersections with flexible and integral traffic control strategies. Second, efficient and scalable optimization framework is necessary to solve the large-scale traffic control optimization problems. In this paper, we first present a centralized traffic control model based on the emerging dynamic path flows. This new model in essence views the whole target network as one integral piece in which traffic propagates based on traffic flow dynamics, vehicle paths, and traffic control. In light of this centralized traffic control concept, most requirements for the existing traffic control coordination will be dropped in the optimal traffic control operations, such as the common cycle length or offsets. Instead, the optimal traffic control strategy across intersections will be highly adaptive over time to minimize the total travel time and it can also prevent any short-term traffic congestions according to the space-time characteristics of vehicle path flows. A mixed integer linear programming (MILP) formulation is then presented to model the propagations of path flows given a centralized traffic control strategy. Secondly, a new approximated Lagrangian decomposition framework is presented. Solving the proposed MILP model with the traditional Lagrangian decomposition approach faces the leveraging challenge between the problem tractability after decomposition and computing complexity. To address this issue, we propose an approximate Lagrangian decomposition framework in which reformulate the target problem is reformulated into an approximated problem by constructing its complicating attributes into the objective function instead of in constraints to reduce the computing complexity. The computing efficiency loss of the objective function can be mitigated via multiple advanced computing techniques. With the proposed approximated approach, the upper bound and tight lower bound can be obtained via two customized dynamic network loading models. This new approach can be generalized to solve many other transportation optimization problems. In the end, one demonstrative and one real-world example are illustrated to show the validness, robustness, and scalability of the new approach.

# 1. Introduction

At the dawn of mobile computing era, the mobile computing devices, from smartphones to connected automated vehicles (CAV) become ubiquitous and contributing novel traffic data sets like vehicle path flows. For instance, some Internet companies are publishing such data sets for research and applications (DiDi Inc., 2017). The path flow data not only reveals vehicles' "presence" at fixed spots but also report vehicles' paths or tours. These new features are substantially pushing the ceilings of both traffic control theories and practices to develop next generation traffic control systems with better intelligence and robustness. In the meantime, the computing powers necessary for large-scale traffic control optimizations is increasingly becoming accessible due to the advancement of high-performance computing (HPC) techniques. Embracing the powers of the emerging traffic data sets and the HPC techniques enables us to explore a revolutionary traffic control optimization framework. This paper is inspired by these emerging potentials and explore a centralized adaptive network traffic control system.

The prevailing traffic control coordination design contains two sequential steps: (1) traffic control optimization according to the local traffic counts at individual intersections and (2) coordinate traffic progression by adjusting offsets. Arguably, such a two-step distributed procedure is based on the distributed optimization and may not be able to capture all the space-time characteristics of traffic propagations, especially when queue spillbacks frequently occur under congested traffic. Based on the emerging path flows, we propose a new centralized adaptive traffic control model to simultaneously optimize traffic control strategies for all intersections. This new model in essence views the whole network as one integral piece within which traffic propagates according to path flows and given dynamic traffic control plans. In light of this centralized traffic control concept, requirements for the traditional traffic control coordination like cycle and offset are no longer necessary. Instead, the optimal timing plan across intersections will be highly adaptive at each intersection to jointly minimize the total travel time. Furthermore, short-term traffic congestions can also be effectively prevented according to the space-time characteristics of vehicle path flows.

There are methodological gaps though to model such a centralized adaptive traffic control systems because (a) the classic traffic control delay models are mostly for individual intersections; (b) full-spectrum traffic propagations among intersections have not been perfectly modelled, especially under congested traffic; and (c) new optimization framework will be necessary to capture the space-time nature of this problem and take advantages of the advanced computing techniques. This paper aims to address the above issues.

## 1.1 Network traffic control based on generalized phase-time network

Li et al. (2015) present a linear traffic control model for individual intersections, namely the phase-time network model. Compared with the other traffic control models, the phase-time network model is able to pre-build most hard constraints inherent in traffic control mechanism into the objective function and maintain the problem's pure linearity, significantly reducing the computing complexity. In this paper, we first enhance the original phase-time network model by explicitly modeling the yellow and all-red clearance between phases rather than implicitly, explicitly modeling permissive phases; adding the "green rest" features to the phase-time network to make sure green stays on the mainline when there is no demand on the side streets and; formulating the link storage constraints for over congested traffic conditions. Compared with the original phase-time network, these improvements make the phase-time network models suitable for all realistic traffic scenarios.

We further generalize the phase-time network model to multiple intersections and serve as the foundation of the proposed centralized adaptive traffic control systems. At any time, there is one and only one local phase activated at an individual intersection. The centralized traffic control status $p_t$ across all intersections at $t$ can be symbolically defined as the combination of all active local phases at individual intersections, $p_t^i$ ($i = 1, 2, \dots n$), as $p_t = \{p_t^1, p_t^2, p_t^3, \dots, p_t^i\}$. Fig. 1 shows a 3-intersection arterial with eight generalized phases. On the right side, Fig. 1 shows a representation of a traffic control coordination with a sequence of generalized phases.

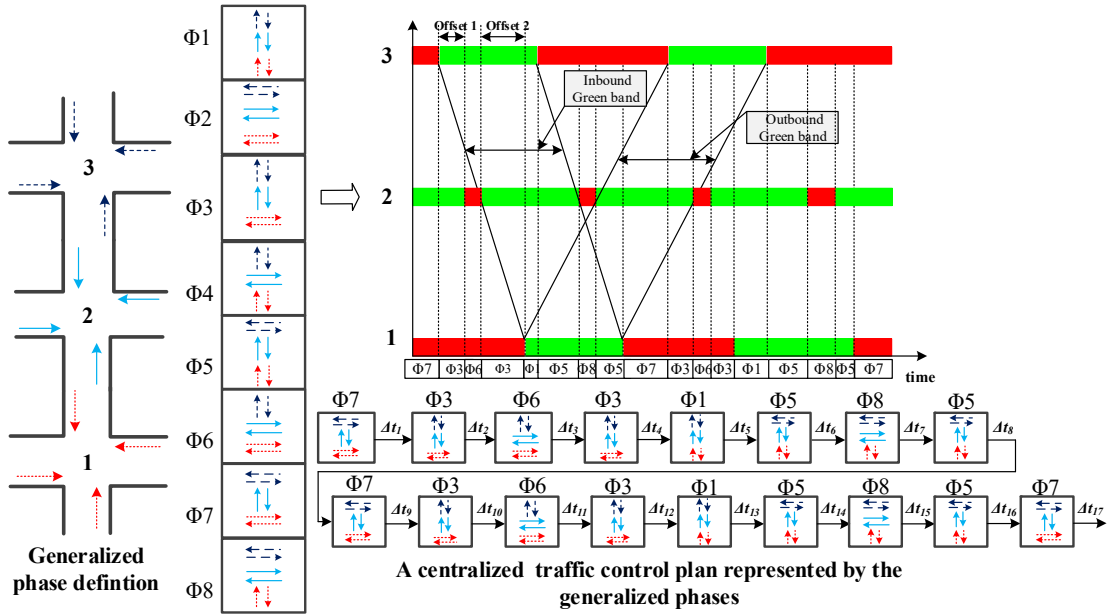

Figure 1 representing the traffic control coordination with the generalized phases

Fig. 1 illustrates the core ideas of the generalized phase-time network model and centralized traffic control mechanism. To performance a centralized traffic control optimization, one only needs to schedule the sequence and durations of generalized phases over time. This is a variant of machine scheduling problems with a rich body of efficient solution algorithms. Any centralized traffic control plan can also be easily translated back to the local traffic control plans at individual intersections.

Nonetheless, with the number of intersections increasing, the number of generalized phases will increase combinatorically and so the optimization will face the "*Curse of Dimensionality*". There are two feasible cures for this issue. One cure is to fix the sequence and durations of certain generalized phases in order to pre-guarantee certain "green bands". For instance, in Fig. 1, one can fix the sequence and durations of $\Phi_3 \rightarrow \Phi_6 \rightarrow \Phi_3 \rightarrow \Phi_1 \rightarrow \Phi_5$ to pre-guarantee the inbound bandwidth between Intersection 2 and 3. Whenever a search algorithm visits $\Phi_3$, the integral "phase group" will be implemented until $\Phi_5$ ends instead of examining any feasible phase next to $\Phi_3$ . Fig. 2 further demonstrates this concept in the generalized phase-time network. This method can significantly reduce the search space. The other cure is to parallelize the search algorithm and/or perform a heuristic search to maintain the search efficiency for various scales of optimizations.

The second challenge is that the proposed traffic control model does not require local minimum and maximum greens at individual intersections. As a result, the final optimal solutions may hold a few vehicles on the minor approaches for an excessively long time in order to minimize the total travel time (i.e., system optimum). This phenomenon can be avoided by superimposing special restrictions of local minimum and maximum greens in a customized search algorithm while seeking the optimal traffic control strategy and it will be elaborated in Appendix B.

The remainder of this paper is organized as follows: literature review on network traffic control strategies is summarized. Then a mixed-integer linear programming (MILP) formulation is presented to describe the space-time relationships between the centralized traffic control strategies and the resulting vehicle trajectories. In the third part, we will present a new scalable approximated Lagrangian decomposition approach to leverage the problem tractability and computing complexity in seeking the lower bound and upper bound solutions. In the last section, two numerical experiments are conducted to examine the robustness and effectiveness of the proposed optimization frameworks.

## 2. Literature Review

Network traffic control optimization consists of two components: (a) network traffic dynamics models and; (b) optimization frameworks. Therefore we divide the literature review into two categories.

### 2.1 Traffic dynamics modeling
One of the most well-known network traffic dynamics representations is the cell transmission model (CTM) (Daganzo, 1994, 1995). The road network is divided into cells with lengths equal to the travel distance per time step. The traffic flow is propagated across road cells based on the equilibrium of four elements: outbound capacity and travel demand of the current cell; inbound capacity and storage capacity of the next cell. The CTM model is in essence a discrete approximation of a linear traffic stream model due to Newell (Newell, 1993a, b). Another traffic dynamics model is the storage-and-forward traffic model due to Gazis and Potts (Gazis and Potts, 1963). In the store-and-forward traffic model, traffic will be stored until the next link has storage capacity under congested traffic conditions. More recently, Han et al. present a continuum approximation for traffic signal control mechanism (Han et al., 2014) based on the *Lighthill–Whitham–Richards* shockwave model (Lighthill and Whitham, 1955; Richards, 1956). The authors argue that the discrete nature of traffic signal control operations make the gradient-directed optimization less effective and the continuum approximation will considerably increase the performance of search algorithms.

The second category of traffic dynamics models for traffic progression problems is the mixed integer linear programming (MILP) approach. Little and Morgan first propose the MILP formulation to maximize the traffic progression on arterials through the adjustment of the traffic control plans (Little and Morgan, 1964). Little et al. further formulate the MILP formulation later on, namely *MAXBAND* (Little et al., 1981). The objective of *MAXBAND* is to maximize a bandwidth linearly dependent on the offsets, the continuous variables in the MILP formulation. The offsets are restricted by constraints with respect to some intermediate integer variables. In the *MAXBAND* formulation, local traffic signal timings (cycle, splits) and travel time ranges between intersections are assumed given. Gartner further extends the *MAXBAND* to *MULTIBAND* to allow for various bandwidths between intersections and add additional flexibility (Gartner et al., 1990). The MILP formulations for *MAXBAND* and *MULTIBAND* serves as the foundation of many network traffic control and coordination researches later. As examples, an asymmetrical version of MULTIBAND or "*AM-BAND*" is to separate the inbound and outbound bands into two independent components and therefore the progression line in *MULTIBAND* is no longer needed (Zhang et al., 2015). This modification is suitable for those arterials with imbalanced mainline traffic; a network version. Another extension from the *MULTIBAND* is "*OD-BAND*" to maximize the weighted bands based on the network origin-destination demands (Arsava et al., 2018). A new feature in the *OD-BAND* is that the phasing sequence is optimized as well.

Other than the variants of *MULTIBAND*, different MILP formulations are also explored in other traffic applications. He et al. present a MILP formulation for multimodal traffic control optimization including regular and connected vehicles (He et al., 2014); Yu et al. conduct integrated optimization of both traffic signal control and vehicle trajectories in which the traffic signal optimization subproblem is formulated as a MILP formulation (Yu et al., 2018); He et al. present a platoon-based arterial multimodal signal control optimization framework in which the critical decision variables is the split ratio of the approaching (given) platoons to minimize the total delays (He et al., 2012). To keep the linearity of the target problem, the authors formulate the constraints of traffic signal mandates based on the traffic control precedence graph diagram due to Head et al.(Head et al., 2006). Li et al. present a new vehicle-oriented MILP formulation to model the intersection automation for a future scenario in which the traffic approaching intersections is heterogeneous and composed of human-driving vehicles and connected automated vehicles (Li et al., 2015; Li and Zhou, 2017). Unlike the MAXBAND or its variants, all the above MILP traffic control models are optimized based on isolated intersections.

The third type of traffic dynamic models for traffic control optimization is implicit in some online traffic control optimization systems. RHODES has embedded a point-queue model to represent the dynamic traffic states (Mirchandani and Head, 2001). Vehicle arrivals are detected via upstream detectors, and the departures are determined by the real-time traffic control operations and result in various levels of delays. Similar traffic dynamics representations are adopted by OPAC (Gartner, 1983), SURTRAC (Xie et al., 2012) in determining the real-time optimal traffic control strategies. These real-time traffic dynamics models are similar with online traffic simulators and no mathematical programming formulations are presented in the literature.

The fourth type of traffic dynamics models is based on nonlinear control delay models. A rich body of the literature on online and off-line traffic signal optimization is based on the traffic dynamics models in this category. One famous network traffic control optimization package is the TRANSYT-7 in which the local traffic dynamics at intersections are based on the arrival profile and nonlinear control delay modes while The traffic propagation between intersections are based on the traffic dispersion model due to Richardson (Bell, 1981; Robertson, 1969).

## 2.2 Traffic control optimization

As for the traffic control optimization, they can also be divided into two categories. For traffic control optimization based on rigorous mathematical programming formulations, the traffic control optimizations are mostly conducted with mathematical solvers, such as CPLEX package. This approach is typically for small problems due to the limitations of memory and computing capacities.

For those traffic control optimization problems based on nonlinear, continuous traffic dynamics models, gradient-directed or metaheuristic search algorithms are commonly adopted, such as the hill climbing optimization in TRANSYT-7 and its extended optimization module based on the Genetic Algorithm (Bell, 1981; Park et al., 1999). Network traffic control optimization is also optimized based on the CTM models and Genetic Algorithm (Lo et al., 2001). Another metaheuristic algorithm, the Ant Colony

Optimization, is also applied to the traffic signal coordination problem under over congested traffic condition (Putha et al., 2012).

Another commonly used optimization technique is the search algorithms such as dynamic programming (DP) or Branch-and-bound algorithms. RHODES formulate the on-line traffic control optimization problem as a knapsack problem and solve to reach the global optimum via DP while OPAC and SURTRAC adopt the approximate DP techniques to reach the quasi-optimum efficiently. Li et al. formulate the dynamic traffic control optimization as a shortest path problem and the shortest path can be found using the label-correcting algorithm, a variant of DP (Li et al., 2015). Li et al. develop a sequential branch-and-bound algorithm to optimize the traffic control strategy within a heterogeneous traffic environment (Li and Zhou, 2017). The selected literature is compared as in Table 1.

Table 1 Comparison of selected literature

| Literature | Category | Formulation Type | Objectives | Constraints | Problem Scope | Solution |
|---|---|---|---|---|---|---|
| CTM by Daganzo (1994,1995) | Traffic Dynamics modeling | MILP | Estimate traffic pattern | Cell Equilibrium of demand, capacity and link story | Network | -- |
| SFM by Gazis and Potts (1963) | Traffic Dynamics modeling | MILP | Estimate traffic pattern | Demand and downstream storage balances | Network | -- |
| Han et al. (2014) | Traffic Dynamics modeling | LP | Transform discrete traffic control problem to continuous approximation | Traffic conservation law | Network | M.P. Solver |
| MAXBAND by Little and Morgan (1981) | Traffic control optimization | MILP | Maximize two-way bandwidths | Logic relations between variables and inputs | Mainline of arterials | M.P. Solver |
| MULTIBAND by Gartner et al. (1990) | Traffic control optimization | MLP | Maximize two-way bandwidths for each link segment | Logic relations between variables and inputs | Mainline of arterials | M.P. Solver |
| He et al. (2014) | Traffic control optimization | MILP | Minimize delays for special vehicles | Traffic control mandates with fixed phasing sequence | Individual intersection | M.P. Solver |
| Li et al. (2015) | Traffic control optimization | MILP | Minimize the total delay | Traffic dynamics and traffic control mandates | Network with distributed optimization | Lagrangian Decomposition and DP |
| RHODES by Mirchandani | Traffic control | -- | Minimize the total delay | Traffic signal mandates (min | Individual intersection | DP |

| | | | | green, max green, etc.) | | |
|---|---|---|---|---|---|---|
| and Head (2001) | optimizatio n | | | green, max green, etc.) | | |
| SURTRAC By Xie et al. (2012) | Traffic control optimization | -- | Scheduling for the coming platoons | Traffic signal mandates (min green, max green, etc.) | Individual intersection | approximate DP |
| OPAC by Gartner (1983) | Traffic control optimization | -- | Minimize the total delay | Traffic signal mandates (min green, max green, etc.) | Individual intersection | approximate DP |
| Wang et al. (This paper) | Traffic control optimization | MILP | Minimize the total delay | Traffic flow constraints and traffic control constraints | Network with centralized optimization | Approximated Lagrangian Decomposition and DP |

## 3. Phase-time network construction for multiple intersections at network traffic control problems

A complete traffic control plan for individual intersections contains three elements: (a) traffic control phases, positive integers indicating one movement or a group of concurrent movements within intersections; (b) phase durations and; (c) phasing sequence. Also, there are two types of phase definitions all over the world: "NEMA Phase" referring to a single movement within intersections in North America and "Stage" referring to a group of concurrent movements in Europe. Li et al. (2015) mathematically prove these two phase definitions are interchangeable in representing traffic control strategies. In the original phase-time network for individual intersections, a phase refers to a group of concurrent vehicle movements within an intersection (i.e., stage). The concept of phases is generalized in this paper in order to represent the centralized network traffic control strategy. A generalized phase is constructed by selecting one and only one phase (stage) at each involved intersection. Only one generalized phase within the traffic network is active at any time to guarantee only concurrent movements is allowed at an individual intersection. This rule will guarantee the traffic safety at all intersections. In addition, an improvement is made to the phase-time network model and the MILP formulation to represent widely existing permissive phases in the real world.

Generating generalized phases and basic configurations across intersections from the local phases at individual sections is described in Algorithm One. Since the concept of generalized phase share similarities with the original concept, we will still use "phase" to represent the generalized phase in the rest of this paper unless confusion is caused.

Algorithm One: Generation of all generalized phases based on the local phases at intersections

Denote $M$ the number of intersections; $N_i$ the number of phases at intersection $i$ ($i \sim [1, M]$); $n_i$ a phase at intersection $i$ ($n \sim [1, N_i], i \sim [1, M]$); $g_{min}(n_i), g_{max}(n_i)$ minimum and maximum greens of $n_i$ respectively; $y(n_i), all\_red(n_i)$ yellow and all-red clearance of $n_i$; $\phi_{<n_1,n_2,n_3,...,n_M>}$ a generalized phase, indexed by a vector $< n_1, n_2, n_3, ..., n_M >$, $LIST$ the list of generalized phases.

For each $n_1$ at intersection 1
    For each $n_2$ at intersection 2
       …
       For each $n_M$ at intersection $M$
          Generate a generalized phase $\phi_{<n_1,n_2,n_3,...,n_M>}$ ;
          Add $\phi_{<n_1,n_2,n_3,...,n_M>}$ into $LIST$;
          $g_{min}(\phi_{<n_1,n_2,n_3,...,n_M>}) = min(g_{min}(n_1), g_{min}(n_2) ..., g_{min}(n_M))$;
          $g_{max}(\phi_{<n_1,n_2,n_3,...,n_M>}) = min(g_{max}(n_1), g_{max}(n_2) ..., g_{max}(n_M))$;
          $y(\phi_{<n_1,n_2,n_3,...,n_M>}) = max(y(n_1), y(n_2) ..., y(n_M))$;
          $all\_red(\phi_{<n_1,n_2,n_3,...,n_M>}) = max(all\_red(n_1), all\_red(n_2) ..., all\_red(n_M))$;
       End
    End
End

The total number of complete generalized phases after Algorithm One is $\prod_{i=1}^{i=M} N_i$ and Fig. 2 demonstrates a possible approach to the search space reduction in the generalized phase-time network to avoid the "Curse of Dimensionality". In a demonstrative one-way arterial with two intersections, the total number of generalized phases are four and the generalized phase-time network can be constructed similarly with the original phase-time network for individual intersections. Nonetheless, we can eliminate a significant amount of phase-time arcs starting from Φ4. Assuming the travel time from Intersection 2 to Intersection 1 is 3 s. Then the reserved phase-time arcs starting from Φ4 in Fig. 2 will always guarantee a green band (i.e., whenever a NB queue is released from Intersection 2 and the queue will always meet a green onset when arrive at the downstream intersection). Compared with the fully populated phase-time networks, the approach to "pre-guaranteeing" some green bands will reduce the search space. The centralized traffic control optimization is to seek the least-cost path from origin to the super sink node $z$ within the phase-time network.

Compared with the original phase-time network for individual intersections, the generalized phase-time network model contains fundamental differences. First, a generalized phase covers all intersections and so evaluating the cost of a generalized phase (i.e., phase-time arc cost) must cover all vehicles within the scope of the road network. It is necessary to track the changes of all vehicles' cumulative path delays given a phase operation. A vehicle's path delay at time $t$ is defined as the difference between $t$ and ideal free-flow travel time arriving at the same location. By contrast, the original phase-time edge costs can be estimated just based on the vehicles' local delays at individual intersections. Second, unlike the original phase-time network, it is difficult to conduct an exact search for the global optimal solution to network traffic signal operations due to the "curse of dimensionality" and so substantial efforts of solving real-world problems must be dedicated to the advanced computing technique to increase the computing efficiency.

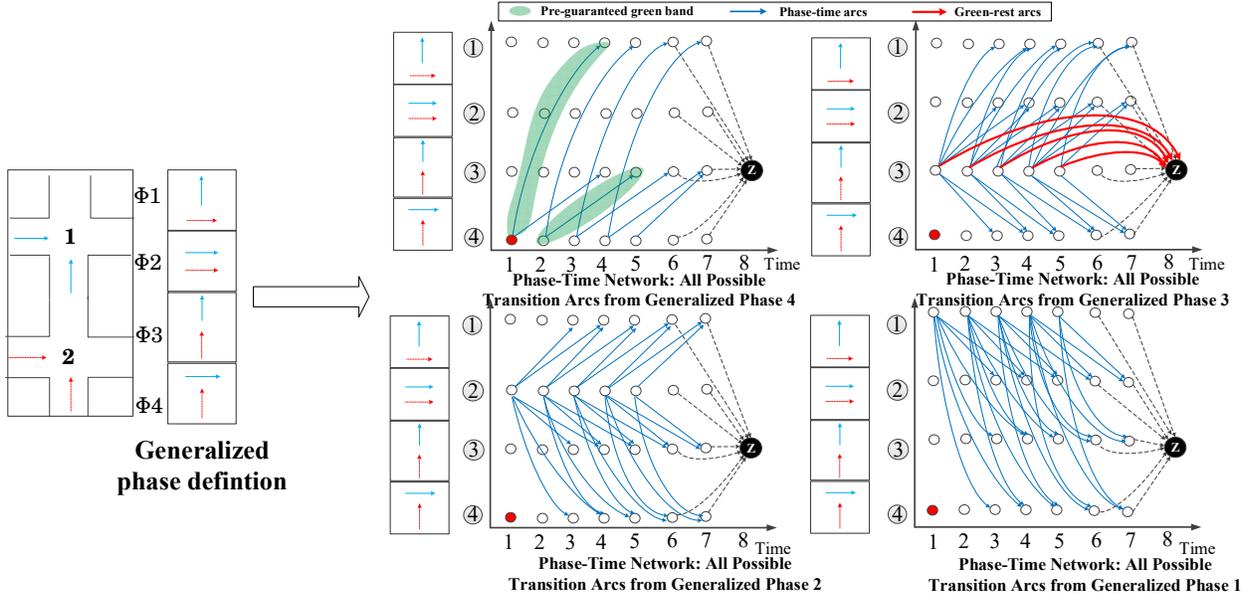

Figure 2 constructing a generalized phase-time network with pre-guaranteed green bands

## 4. Mixed integer linear programming (MILP) formulation for centralized traffic control optimization

In this section, we present a MILP formulation to optimize the centralized traffic control strategy. It serves as the methodological foundation of the solution algorithms for the target problem.

Table 2 Parameters and variables for the MILP formulation

| **Notations for road network and space-time network** | |
|---|---|
| $G_o(N_o, A_o)$ | Road network |
| $G(N, A)$ | Space-time network |
| $N_o, A_o, A_s$ | $N_o$: set of road network nodes; $A_o$: set of road network links; $A_s$: road links controlled by traffic lights (open if the corresponding phase is green (active); red (inactive) otherwise) |
| $N, A$ | $N$: set of space-time network nodes; $A$: set of space-time network arcs |
| $i, j, (i, j)$ | $i, j \in N_o, (i, j) \in A_o$ |
| $t, s, \tau, h, \mathcal{H}$ | Time indices; $\mathcal{H}$: time horizon |
| $t_0^v$ | Departure time of $v$ |
| $FFTT_{(i,j)}$ | Free flow travel time on link $(i, j), \forall (i, j) \in A_o$ |
| $SR_{(i,j)}$ | Saturation rate of $(i, j), \forall (i, j) \in A_o$ |
| $L(i, j)$ | The storage capacity of $(i, j) \in A_o$ |
| $(i, t), (j, s), (i, t, j, s)$ | $(i, t), (j, s) \in N, \forall (i, t, j, s) \in A; \ if \ (i, j) \in A_o, s = t + FFTT_{(i,j)}; \ if \ i = j$ (waiting arc), $s = t + 1$ |
| $v, V$ | Vehicle $v \in V$ |
| $o(v), d(v)$ | Origin and destination of $v$ |
| $\varphi_{(v,i,j)}, \Phi$ | $\varphi_{(v,i,j)} = 1$ if $(i,j)$ is the last link of $v$'s path; 0 otherwise; $\Phi = \{\varphi_{(v,i,j)}\}, \forall v \in V, \forall (i,j) \in A_0$ |
| $\omega_{(v,i,j)}, \Omega$ | $\omega_{(v,i,j)} = \omega_{(v,i,i)} = \omega_{(v,j,j)} = 1$, if $(i,j)$ is on the $v$'s path; 0 otherwise; $\Omega = \{\omega_{(v,i,j)}\}, \forall v \in V, \forall (i,j) \in A_0$ |
| $c_{(v,i,j)}, \mathbb{C}$ | $c_{(v,i,j)}$: total free-flow path travel time if $(i,j)$ is the last link of $v$'s path; otherwise 0; $\mathbb{C} = \{c_{(v,i,j)}\}, \forall v \in V$ |
| $e_{(v,i,j)}$ | $e_{(v,i,j)}$: $v$'s earliest (non-delay) time to enter $(i,j)$ if $\omega_{(v,i,j)} = 1$; otherwise 0 |
| **Notations for phase-time network** | |
| $\Psi(P, T)$ | Generalized phase-time network |
| $P$ | Set of generalized phases |
| $p_i^l$ | Local phase at intersection $i$ |
| $(p, t), (p', t')$ | Vertices in $\Psi$, $p, p' \in P, t, t' \in T$. Symbolically, $p$ is a composition of one and only one local phase at each intersection $i$. $p = \{p_1^l, p_2^l, \dots, p_i^l\}$ |
| $(p_o, 0), (p_z, H)$ | Origin (current phase) and destination vertex (ending phase) in $\Psi$ |
| $(p, t, p', t')$ | A phase-time edge in $\Psi$, representing "phase $p$ starts green at $t$, after yellow and all-red, turns over green to phase $p'$ at $t'$"; note $p \neq p'$ because there are no waiting arcs in the phase time networks |
| $y_p, ar_p, \rho_p^g, \rho_p^{ar}, \rho_p^y$ | $y_p, ar_p$: The yellow and all-red times for $p$; $\rho_p^g, \rho_p^y, \rho_p^{ar}$: the link capacity discount factor during green, yellow and all-red, $\rho_p^g = 1, \rho_p^y \sim (0,1), \rho_p^{ar} = 0$; |
| $\delta_p$ | Signal link capacity discount factor for permissive phase |
| $m_{(i,j,p)}, \mathcal{M}$ | $m_{(i,j,p)} = 1$ if $(i,j)$ is controlled by a protected phase $p$; , $\delta$ if $(i,j)$ is controlled by a permissive phase $p$; and 0 otherwise. $\mathcal{M} = \{m_{(i,j,p)}\}; \forall (i,j) \in A_s, \forall p \in P$ |
| $f(v, v', i, j), \mathcal{F}$ | Vehicle first-in-first-out (FIFO) relationship matrix, $f(v, v', i, j) = 1$ if $v$ must enter link $(i,j)$ earlier than $v'$; otherwise 0. $v, v' \in V, (i,j) \in A_o, \mathcal{F} = \{f(v, v', i, j)\}$ |
| $Q$ | A very large positive integer |
| **Variables** | |
| $x_{(v,i,t,j,s)} \in X$ | Equal to 1 if $v$ enters link $(i,j)$ at $t$ and leaves at $s$; otherwise 0 |
| $y_{(p,\tau,p',h)} \in Y$ | equal to 1 if and only if the phase-time arc $(p, \tau, p', h)$ is, otherwise 0 |
| $\gamma_{(i,j,t)}$ | Open-close status indicator of $(i,j), \forall (i,j) \in A_o$. For $\forall (i,j) \in A_s$, $\gamma_{(i,j,t)} = 1$ if $(i,j)$ is open at $t$, 0 otherwise; for $\forall (i,j) \in \{A_o - A_s\}$, always equal to 1 |
| $\lambda_i(i, j, t, s)$ | Lagrangian multipliers for the relaxed constraints $(i)$ |

$$\min Z_1 = \sum_{v \in V} \sum_{s \leq \mathcal{H}} \sum_{1 \leq t} \sum_{(i,j) \in A_o} \left( (s - t_o^v) \times \varphi_{(v,i,j)} \times x_{(v,i,t,j,s)} \right) - \sum_{v \in V} \sum_{(i,j) \in A_o} c_{(v,i,j)} + \sum_{(p,\tau,p',h) \in \Psi} y_{(p,\tau,p',h)}$$

$$(1)$$

The third term on the right side of (1) is to prevent unnecessary phase transitions if there are no vehicles on the side streets (a.k.a. the "Green Rest").

Subject to:

<u>Road link capacity constraints in $G(N, A)$</u>

For road links controlled by infrastructure like traffic signal systems:

$$\sum_{v \in V} \omega_{(v,i,j)} \times x_{(v,i,t,j,s)} \leq \gamma_{(i,j,t)} \times SR_{(i,j)}, t, s \sim [1, \mathcal{H}], \forall (i,j) \in A_s \tag{2}$$

Extended from the approach due to Li et al. (2015), constraints (2) can be reformulated with an additive linear summation of binary phase-time arc variable $y_{(p,\tau,p',h)}$ as:

$$\sum_{v \in V} \omega_{(v,i,j)} \times x_{(v,i,t,j,s)} \leq \Big( \sum_{p \neq p', p \in P} \sum_{p' \in P} \sum_{t+y_p+ar_p < h \leq \mathcal{H}} \sum_{1 \leq \tau \leq t} \Big( \rho_p^g \times m_{(i,j,p)} \times y_{(p,\tau,p',h)} \Big) +$$

$$\sum_{p \neq p', p \in P} \sum_{p' \in P} \sum_{t+ar_p < h \leq t+y_p+ar_p} \sum_{1 \leq \tau \leq t} \Big( \rho_p^y \times m_{(i,j,p)} \times y_{(p,\tau,p',h)} \Big) +$$

$$\sum_{p \neq p', p \in P} \sum_{p' \in P} \sum_{t < h \leq t+ar_p} \sum_{1 \leq \tau \leq t} \Big( \rho_p^{ar} \times m_{(i,j,p)} \times y_{(p,\tau,p',h)} \Big) \Big) \times SR_{(i,j)}, t, s \sim [1, \mathcal{H}], \forall (i,j) \in A_s \tag{3}$$

Whenever a generalized transition takes place, at least one local phase must enter yellow and all-red therefore constraints (3) separate the capacity constraints on control links into green, yellow and all-red clearance terms. Fig. 3 demonstrates the boundaries of green, yellow and all-red clearance terms on the right side of (3) as opposed to $t$ for a phase-time arc $(p, \tau, p', h)$ to determine the traffic signal status and control link capacity at any $t$.

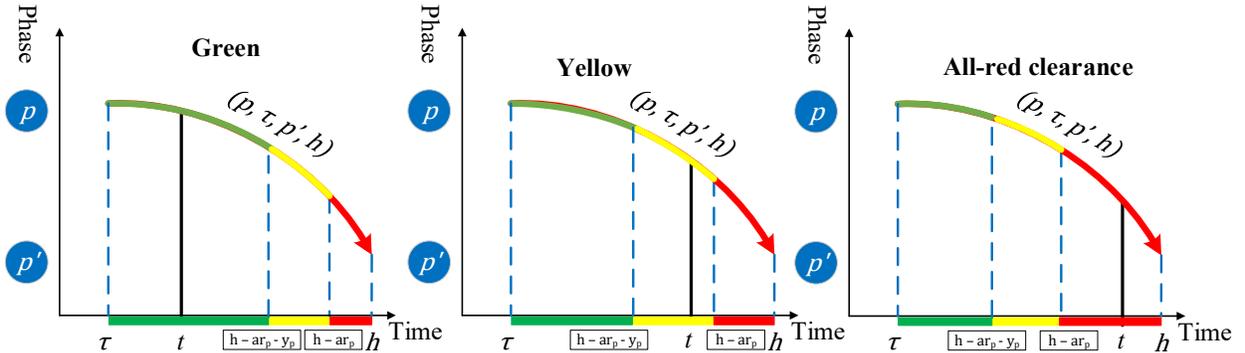

Figure 3 Status boundaries of control links at $t$

According to the concept of the generalized phase-time network, whenever a generalized phase transition takes place $(p \rightarrow p')$, all local phases must simultaneously end at each intersection for safety reasons. However, such a transition rule may generate less favored control operations at some intersections in that, if a local phase $p_i^l$ at Intersection $i$ happens to be part of both $p$ and next $p'$, $p_i^l$ should stay green instead of entering yellow and all-red and turn back to green. To address this issue, (3) are further modified as:

$$\sum_{v \in V} \omega_{(v,i,j)} \times x_{(v,i,t,j,s)} \leq \Big( \sum_{p \neq p', p \in P} \sum_{p' \in P} \sum_{t+y_p+ar_p < h \leq H} \sum_{1 \leq \tau \leq t} \Big( \rho_p^g \times m_{(i,j,p)} \times y_{(p,\tau,p',h)} \Big) +$$

$$\sum_{p \neq p', p \in P} \sum_{p' \in P} \sum_{t+ar_p < h \leq t+y_p+ar_p} \sum_{1 \leq \tau \leq t} \Big( \rho_p^y \times m_{(i,j,p)} \times \big( 1 - m_{(i,j,p')} \big) \times y_{(p,\tau,p',h)} \Big) +$$

$$\sum_{p \neq p', p \in P} \sum_{p' \in P} \sum_{t+ar_p < h \leq t+y_p+ar_p} \sum_{1 \leq \tau \leq t} \Big( \rho_p^g \times m_{(i,j,p')} \times m_{(i,j,p')} \times y_{(p,\tau,p',h)} \Big) +$$

$\sum_{p \neq p', p \in P} \sum_{p' \in P} \sum_{t < h \leq t + ar_p} \sum_{1 \leq \tau \leq t} \left( \rho_p^{ar} \times m_{(i,j,p)} \times \left( 1 - m_{(i,j,p')} \right) \times m_{(i,j,p')} \times y_{(p,\tau,p',h)} \right) +$

$\sum_{p \neq p', p \in P} \sum_{p' \in P} \sum_{t < h \leq t + ar_p} \sum_{1 \leq \tau \leq t} \left( \rho_p^g \times m_{(i,j,p)} \times m_{(i,j,p')} \times y_{(p,\tau,p',h)} \right) \right) \times SR_{(i,j)}, t, s \sim [1, H], \forall (i,j) \in A_s$   (3')

In (3'), the yellow term on the right side is expanded into two exclusive terms. If a signal link $(i,j) \in A_s$ is controlled by two consecutive phases $p, p'$ (i.e., $m_{(i,j,p)} = m_{(i,j,p')} = 1$) , then

$\sum_{p \neq p', p \in P} \sum_{p' \in P} \sum_{t + ar_p < h \leq t + y_p + ar_p} \sum_{1 \leq \tau \leq t} \left( \rho_p^g \times m_{(i,j,p')} \times m_{(i,j,p')} \times y_{(p,\tau,p',h)} \right)$ (the 3rd term on the right side) takes effects like a continuous long green; otherwise $m_{(i,j,p)} = 1, m_{(i,j,p')} = 0$ and

$\sum_{p \neq p', p \in P} \sum_{p' \in P} \sum_{t + ar_p < h \leq t + y_p + ar_p} \sum_{1 \leq \tau \leq t} \left( \rho_p^y \times m_{(i,j,p)} \times \left( 1 - m_{(i,j,p')} \right) \times y_{(p,\tau,p',h)} \right)$ (the 2nd term on the right side) take effects as normal yellow. The same analysis applies to all-red clearance terms.

For regular road links:

$\sum_{v \in V} \omega_{(v,i,j)} \times x_{(v,i,t,j,s)} \leq SR_{(i,j)}, t, s \sim [1, \mathcal{H}], \forall (i,j) \in (A_o - A_s)$   (4)

Road link storage constraints in $G(N, A)$

Constraints (5) guarantee the number of vehicles on a link at any time will not exceed the maximal link storage.

$\left( \sum_{0 \leq \tau \leq t} \sum_{v \in V} \left( \omega_{(v,i,j)} \times x_{(v,i,\tau,j,\tau + FFTT_{(i,j)})} \right) \right) - \left( \sum_{0 \leq \tau \leq t} \sum_{v \in V} \sum_{(j,\tau,i,\tau + FFTT_{(j,i)}) \in A} \left( \omega_{(v,j,i)} \times x_{(v,j,\tau,i,\tau + FFTT_{(i,j)})} \right) \right) \leq L_{(i,j)}, \forall (i,j), (j,i) \in A_o, t \sim [1, \mathcal{H}]$   (5)

The first term on the left side of (5) represents the cumulative vehicles arriving at i at t (i.e., the "A curve") while the second term represents the cumulative vehicles departing j at t.The difference is the number of vehicles on $(i,j)$ at t (i.e., the "D curve").

First-In-First-Out (FIFO) constraints in $G_o(N_o, A_o)$

Like most analytic formulations of traffic network modeling and dynamic traffic assignment, we assume the FIFO rule for vehicles crossing the same link. If a vehicle departs earlier than another vehicle and they are on the same path, then the first vehicle with early departure time will enter all the links along the path before the second vehicle. In the context of this paper, all vehicles' paths and departure times are assumed known as inputs and their FIFO relationship is defined in $\mathcal{F}$.

$\sum_{t = 1,2,...\mathcal{H}} \left( t \times x(v, i, t, j, t + FFTT(i,j)) \right) \leq \sum_{t = 1,2,...\mathcal{H}} \left( t \times x(v', i, t, j, t + FFTT(i,j)) \right), \forall v, v' \in V, (i,j) \in A_o, f(v, v', i, j) = 1$   (6)

Flow conservation constraints in space-time network $G(N, A)$

The space-time flow conservation constraints guarantee vehicles to move from their origins to destinations along their paths. Constraints (7) hold:

$$\sum_{(i,t,j,s)\in A}\left(\omega_{(v,i,j)}\times x_{(v,i,t,j,s)}\right)-\sum_{(j,s,i,s')\in A}\left(\omega_{(v,j,i)}\times x_{(v,j,s,i,s')}\right)=\begin{cases}-1;\ (j,s)=o_v\\ \ \ 1;\ (j,s)=d_v\\ \ \ 0;\ otherwise\end{cases},\forall v\in V,\forall(j,s)\in N \qquad (7)$$

Flow conservation constraints phase-time network $\Psi(P,T)$

A feasible traffic signal plan can be represented by a path in $\Psi(P,T)$ from the origin vertex $(p_o,0)$ to the super sink vertex. $(p_z,T)$. Therefore the flow conservation constraints are needed as in (8).

$$\sum_{(p,\tau,p',h)\in\Psi}y_{(p,\tau,p',h)}-\sum_{(p',h,p,h')\in\Psi}y_{(p',h,p,\ h')}=\begin{cases}-1;\ (p',h)=(p_o,0),\\ \ \ 1;\ (p',h)=(p_z,H),\\ \ \ \ 0;\ otherwise\end{cases}for\ \forall(p',h)\in N \qquad (8)$$

**Remark:** $\mathcal{M}$ in $Z_1$ is *0- δ -1* matrix to map signal-controlled links and traffic signal phases. The 0-1 elements reflect the protected phases (for the exclusive right of ways). In reality, there are also permissive phases. A permissive green means that vehicles are allowed to cross the intersection if they can find sufficient gaps in opposing traffic. In other words, some elements in $\mathcal{M}$ can be a positive value between 0 and 1 to reflect the reduced link capacity during the permissive phases in (3').

# 6. Approximated Lagrangian decomposition Solution

$Z_1$ is a space-time network optimization problem and contains enormous variables and constraints. In order to solve a large-scale space-time network problem, most relevant literature adopts the Lagrangian decomposition approach. The key to the success of this approach is how to design an efficient relaxation problem (i.e., the lower bound problem) to decompose and then solve. The proposed MILP formulation in this paper contains many complicating constraints and therefore we face a challenge of designing an appropriate relaxation problem efficiently decomposable and solvable. To address this issue, we propose a new approximated Lagrangian decomposition solution to leverage the solutions between mathematical rigorousness and solution efficiency.

Traditional formulations for optimization problems typically keep the succinctness of the objective function and model the complicating attributes into constraints. In the proposed approximated Lagrangian decomposition solution, we reformulate many complicating attributes into the objective function. As a result, the original MILP formulation can be transformed into an approximated problem based on the dynamic-network-loading (DNL) type of objective functions. Fig. 4 shows how to reformulate the target problem.

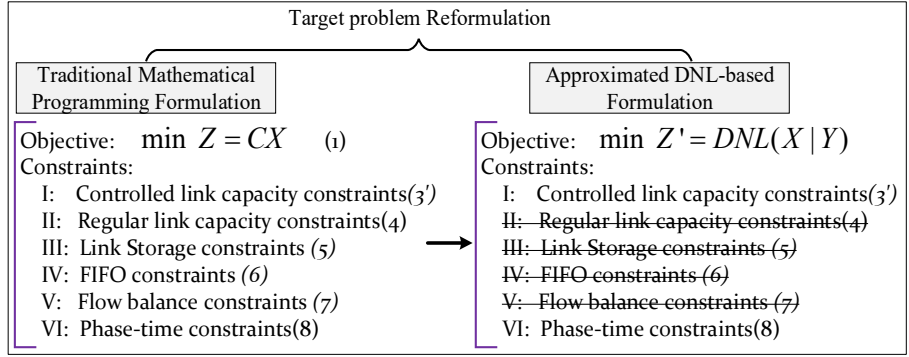

Figure 4 Transforming from a M.P. formulation to the DNL-based formulation

Relocating the problem's complicating attributes from the constraints into the objective function can greatly reduce the computing complexity. The loss of efficiency on the side of objective function can also be mitigated with multiple high performance computing techniques for the DNL expedition.

Fig. 5 shows the framework of the proposed approximated Lagrangian decomposition solutions. The core parts of this framework is to develop two DNL models: standard DNL for feasible solutions (upper bound) and customized DNL for part of the lower bound.

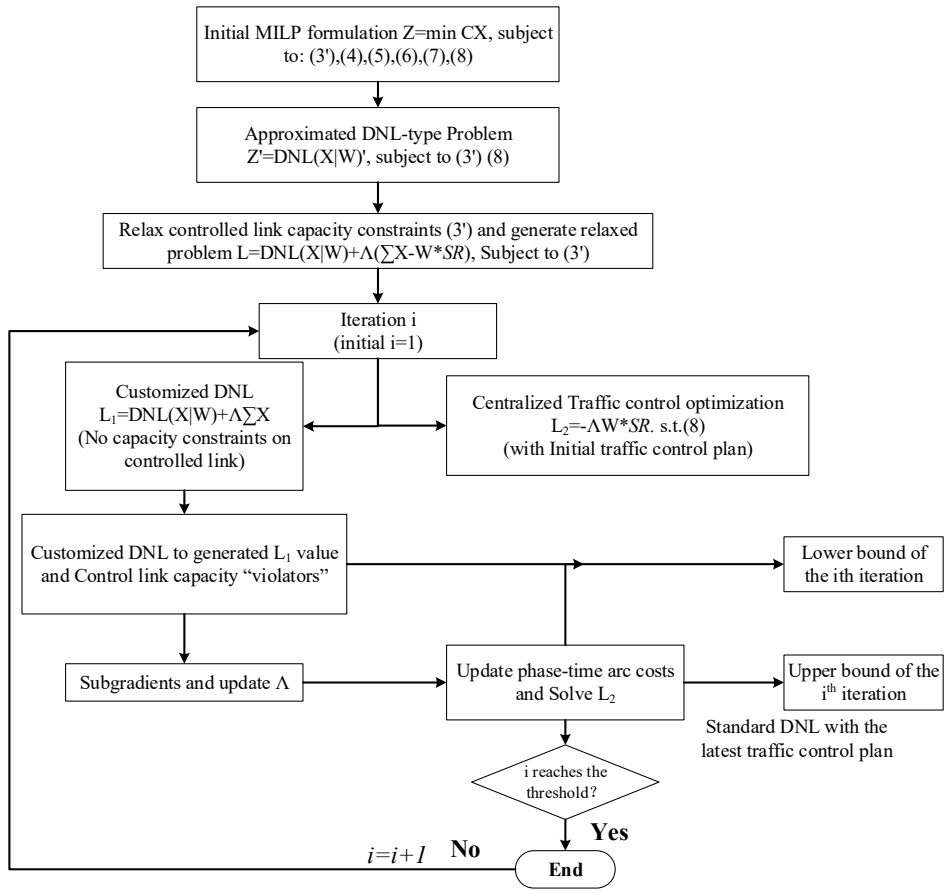

Figure 5 Framework of approximated Lagrangian decomposition approach

6.1 Lagrangian decomposition solution

Dualizing constraints (3') will create a Lagrangian relaxation problem $\mathcal{L}_1$ which is a lower bound of $Z_1$ and its formulation is as follows:

$$\min \mathcal{L}_1 = \left(\sum_{v \in V} \sum_{s \le \mathcal{H}} \sum_{1 \le t} \sum_{(i,j) \in A_o} \left(s \times \varphi_{(v,i,j)} \times x_{(v,i,t,j,s)}\right) - \sum_{v \in V} \sum_{(i,j) \in A_o} c_{(v,i,j)}\right) + \sum_{s \le \mathcal{H}} \sum_{1 \le t} \sum_{(i,j) \in A_s} \Big(\lambda_3(i,j,t) \times$$

$$\left(\sum_{v \in V} \omega_{(v,i,j)} \times x_{(v,i,t,j,s)} - \left(\sum_{p \ne p', p \in P} \sum_{p' \in P} \sum_{t+y_p+ar_p < h \le \mathcal{H}} \sum_{1 \le \tau \le t} (\rho_p^g \times m_{(i,j,p)} \times y_{(p,\tau,p',h)}) + \right.\right.$$

$$\sum_{p \ne p', p \in P} \sum_{p' \in P} \sum_{t+ar_p < h \le t+y_p+ar_p} \sum_{1 \le \tau \le t} \left(\left(\rho_p^y \times m_{(i,j,p)} \times \left(1 - m_{(i,j,p')}\right) + \rho_p^g \times m_{(i,j,p')} \times m_{(i,j,p')}\right) \times y_{(p,\tau,p',h)}\right) +$$

$$\sum_{p \ne p', p \in P} \sum_{p' \in P} \sum_{t < h \le t+ar_p} \sum_{1 \le \tau \le t} \left(\left(\rho_p^{ar} \times m_{(i,j,p)} \times \left(1 - m_{(i,j,p')}\right) + \rho_p^g \times m_{(i,j,p')} \times m_{(i,j,p')}\right) \times y_{(p,\tau,p',h)}\right)\right) \times SR_{(i,j)}\right) \quad (9)$$

Subject to (4), (5), (6) and (7)

The constant $-\left(\sum_{v \in V} \sum_{(i,j) \in A_o} c_{(v,i,j)}\right)$ in (9) can be dropped and (9) is re-organized as (10).

$$\text{Min } \mathcal{L}_1 = \sum_{v \in V} \sum_{s \le \mathcal{H}} \sum_{1 \le t} \left(\sum_{(i,j) \in A_o} \left(s \times \varphi_{(v,i,j)} \times x_{(v,i,t,j,s)}\right) + \sum_{(i,j) \in A_s} \lambda_3(i,j,t) \times \left(\omega_{(v,i,j)} \times x_{(v,i,t,j,s)}\right)\right) +$$

$$\sum_{s \le \mathcal{H}} \sum_{1 \le t} \sum_{(i,j) \in A_s} \left(-\lambda_3(i,j,t) \times \left(\sum_{p \ne p', p \in P} \sum_{p' \in P} \sum_{t+y_p+ar_p < h \le \mathcal{H}} \sum_{1 \le \tau \le t} (\rho_p^g \times m_{(i,j,p)} \times y_{(p,\tau,p',h)}) + \right.\right.$$

$$\sum_{p \ne p', p \in P} \sum_{p' \in P} \sum_{t+ar_p < h \le t+y_p+ar_p} \sum_{1 \le \tau \le t} \left(\left(\rho_p^y \times m_{(i,j,p)} \times \left(1 - m_{(i,j,p')}\right) + \rho_p^g \times m_{(i,j,p')} \times m_{(i,j,p')}\right) \times y_{(p,\tau,p',h)}\right) +$$

$$\sum_{p \ne p', p \in P} \sum_{p' \in P} \sum_{t < h \le t+ar_p} \sum_{1 \le \tau \le t} \left(\left(\rho_p^{ar} \times m_{(i,j,p)} \times \left(1 - m_{(i,j,p')}\right) + \rho_p^g \times m_{(i,j,p')} \times m_{(i,j,p')}\right) \times y_{(p,\tau,p',h)}\right)\right) \times SR_{(i,j)}\right) \quad (10)$$

We can also tell that $\mathcal{L}_1$ is a separable problem and can minimize over $x_{(v,i,t,j,s)}$ and $y_{(p,\tau,p',h)}$ as two sub-problems:

$$\text{Min } \mathcal{L}_1^1 = \sum_{v \in V} \sum_{s \le \mathcal{H}} \sum_{1 \le t} \left(\sum_{(i,j) \in A_o} \left(s \times \varphi_{(v,i,j)} \times x_{(v,i,t,j,s)}\right) + \sum_{(i,j) \in A_s} \lambda_3(i,j,t,s) \times \left(\omega_{(v,i,j)} \times x_{(v,i,t,j,s)}\right)\right) \quad (11)$$

Subject to: constraints (4), (5), (6) and (7)

$$\text{Min } \mathcal{L}_1^2 = \sum_{s \le \mathcal{H}} \sum_{1 \le t} \sum_{(i,j) \in A_s} \left(-\lambda_3(i,j,t) \times \left(\sum_{p \ne p', p \in P} \sum_{p' \in P} \sum_{t+y_p+ar_p < h \le \mathcal{H}} \sum_{1 \le \tau \le t} (\rho_p^g \times m_{(i,j,p)} \times y_{(p,\tau,p',h)}) + \right.\right.$$

$$\sum_{p \ne p', p \in P} \sum_{p' \in P} \sum_{t+ar_p < h \le t+y_p+ar_p} \sum_{1 \le \tau \le t} \left(\left(\rho_p^y \times m_{(i,j,p)} \times \left(1 - m_{(i,j,p')}\right) + \rho_p^g \times m_{(i,j,p')} \times m_{(i,j,p')}\right) \times y_{(p,\tau,p',h)}\right) +$$

$$\sum_{p \ne p', p \in P} \sum_{p' \in P} \sum_{t < h \le t+ar_p} \sum_{1 \le \tau \le t} \left(\left(\rho_p^{ar} \times m_{(i,j,p)} \times \left(1 - m_{(i,j,p')}\right) + \rho_p^g \times m_{(i,j,p')} \times m_{(i,j,p')}\right) \times y_{(p,\tau,p',h)}\right)\right) \times SR_{(i,j)}\right) \quad (12)$$

And it can be re-written as:

$$\text{Min } \mathcal{L}_1^2 = \sum_{p \ne p', p \in P} \sum_{p' \in P} \sum_{t+y_p+ar_p < h \le \mathcal{H}} \sum_{1 \le \tau \le t} \left(\left(\sum_{t+FFTT(i,j) \le \mathcal{H}} \sum_{t \ge 1} \sum_{(i,j) \in A_s} (-\lambda_3(i,j,t) \times \rho_p^g \times m_{(i,j,p)} \times SR_{(i,j)})\right) \times \right.$$

$$y_{(p,\tau,p',h)}\right) + \sum_{p \ne p', p \in P} \sum_{p' \in P} \sum_{t+ar_p < h \le t+y_p+ar_p} \sum_{1 \le \tau \le t} \left(\left(\sum_{t+FFTT(i,j) \le H} \sum_{t \ge 1} \sum_{(i,j) \in A_s} (-\lambda_3(i,j,t) \times (\rho_p^y \times m_{(i,j,p)} \times (1 - \right.$$

$$m_{(i,j,p')}) + \rho_p^g \times m_{(i,j,p')} \times m_{(i,j,p')}) \times SR_{(i,j)})\right) \times y_{(p,\tau,p',h)}\right) +$$

$$\sum_{p \ne p', p \in P} \sum_{p' \in P} \sum_{t < h \le t+ar_p} \sum_{1 \le \tau \le t} \left(\left(\sum_{t+FFTT(i,j) \le H} \sum_{t \ge 1} \sum_{(i,j) \in A_s} (-\lambda_3(i,j,t) \times (\rho_p^{ar} \times m_{(i,j,p')} \times (1 - m_{(i,j,p')}) + \rho_p^g \times \right.\right.$$

$$m_{(i,j,p')} \times m_{(i,j,p')})(\rho_p^{ar} \times m_{(i,j,p)} \times (1 - m_{(i,j,p')}) + \rho_p^g \times m_{(i,j,p')} \times m_{(i,j,p')})SR_{(i,j)})\right) \times y_{(p,\tau,p',h)}\right) \quad (12')$$

Subject to: constraints (8)

(12') is the objective function of the shortest path problem in $\Psi(P, T)$ with respect to $y_{(p,\tau,p',h)}$ and the corresponding phase-time arc costs. In realistic applications, if $(i, j) \in A_s$, then $\varphi_{(v,i,j)} = 0$ because the signal links (within intersections) should never be used as the ending link of a vehicle's path. As such $\mathcal{L}_1^1$ could be further simplified without loss of generality as (13):

$$\text{Min } \mathcal{L}_1^1 = \sum_{v \in V} \sum_{s \leq \mathcal{H}} \sum_{1 \leq t} \left( \sum_{(i,j) \in A_o - A_s} \left( s \times \varphi_{(v,i,j)} \times x_{(v,i,t,j,s)} \right) + \sum_{(i,j) \in A_s} \lambda_3(i,j,t) \times \left( \omega_{(v,i,j)} \times x_{(v,i,t,j,s)} \right) \right) \text{ (13)}$$

$\mathcal{L}_1^1$ and $\mathcal{L}_1^2$ are coupled through Lagrangian multipliers and minimize $\mathcal{L}_1$ is equivalent to minimize $(\mathcal{L}_1^1 + \mathcal{L}_1^2)$. In other literature, this approach is often called the *primal decomposition* (Boyd and Vandenberghe, 2004)

<u>Optimization framework</u>

To minimize $\mathcal{L}_1^1$, we can build a customized DNL model, $DNL_c(X)$, to obtain an approximate optimum because (4), (5), (6) and (7) are a basic set of spatial and temporal constraints for a dynamic network loading (DNL) process except for the capacity constraints of those controlled links. The customized DNL model, for $\mathcal{L}_1^1$ is developed as follows:

- For regular links: apply (4), (5), (6) and (7) link capacity, link storage and flow balance constraints
- For control links: only apply (5), (6) and (7) for link storage and flow balance constraints
- When a vehicle enters a controlled link $(i, j) \in A_s$ (i.e., $\omega_{(v,i,j)} = 1$), it will incur an additional Lagrangian price $\lambda_3(i, j, t)$

The reformulated $\mathcal{L}_1^1$ is: $\text{Min } \mathcal{L}_1'^1 = DNL_c(X)$ with no constraints. Obviously, given the (fixed) vehicle paths (entering times and paths) and Lagrangian marginal prices on controlled links in this context, the total travel times will be minimized after the customized DNL model finishes loading.

$\mathcal{L}_1^2$ is a standard formulation for the shortest path problem in $\Psi(P, T)$. The phase-time arc costs are calculated according to the given $\lambda_3$ values and (12'). The shortest path problem can be solved in (pseudo)polynomial time. The found shortest path can also be interpreted as the optimal centralized traffic control strategy for that iteration.

Minimizing $(\mathcal{L}_1^1 + \mathcal{L}_1^2)$ provides a lower bound (LB) of primal problem $Z_1$ which is not necessarily a feasible solution to $Z_1$. To obtain a feasible solution (the upper bound (UB) to $Z_1$ as well), it is necessary to run the standard DNL process again with the traffic control strategy optimized from $\mathcal{L}_1^2$. It is necessary to adjust the Lagrangian multipliers in order to reduce the gap between LB and UB through iterations. We choose the subgradient method as it can achieve the fastest converging rate (Fisher, 1981).

Since constraints (3') are relaxed to create $\mathcal{L}_1$, the subgradients with respect to $\lambda_3(i, j, t, s)$ ($\forall (i, j) \in A_s, t, s \sim [1, \mathcal{H}]$) are calculated as:

$$\nabla L_{\lambda_3(i,j,t)} = \left( \sum_{v \in V} \omega_{(v,i,j)} \times x_{(v,i,t,j,s)} - \sum_{p,p' \in P, p \neq p'} \sum_{t+y_p + ar_p < h \leq H} \sum_{0 \leq \tau \leq t} (\rho_p^g \times m_{(i,j,p)} \times y_{(p,\tau,p',h)}) \times SR_{(i,j)} \right) \quad (14)$$

$$\nabla L_{\lambda_3(i,j,t)} = \left( \sum_{v \in V} \omega_{(v,i,j)} \times x_{(v,i,t,j,s)} - \sum_{p,p' \in P, p \neq p'} \sum_{t+ar_p < h \leq t+y_p + ar_p} \sum_{0 \leq \tau \leq t} (\rho_p^y \times m_{(i,j,p)} \times y_{(p,\tau,p',h)}) \times SR_{(i,j)} \right) \quad (15)$$

$$\nabla L_{\lambda_3(i,j,t)} = \left( \sum_{v \in V} \omega_{(v,i,j)} \times x_{(v,i,t,j,s)} - \sum_{p,p' \in P, p \neq p'} \sum_{t < h \leq t+ar_p} \sum_{0 \leq \tau \leq t} (\rho_p^{ar} \times m_{(i,j,p)} \times y_{(p,\tau,p',h)}) \times SR_{(i,j)} \right) \quad (16)$$

The first term on the right side of (14) (15) (16) is the results after the special DNL process in $\mathcal{L}_1^1$ and the second term is the resulting capacity of controlled links given the optimal network control plan or the shortest path in $\Psi(P, T)$ in $\mathcal{L}_1^2$ in the same iteration.

<u>Interpretation of the Lagrangian multipliers $\Lambda$ from the view of space-time networks</u>

The Lagrangian multipliers in the approximated Lagrangian decomposition solution are the critical bonds between two subproblems. In addition, the underlying rationale for the subgradients can also be well interpreted from the view of space-time networks. In the customized DNL process, the controlled link capacity constraints are ignored (i.e., they become infinitive). As a result, vehicles enter the intersections during red. The more vehicles run the red light, the more the Lagrangian multipliers on the corresponding controlled link (space) at the corresponding time (time) will increase according to the subgradient equations. Since the values of Lagrangian multipliers are negatively relevant with the corresponding phase-time arcs as defined in (12'), the increase of Lagrangian multipliers will decrease the costs of the corresponding phase-time arcs. In turn, those phase-time arcs will be more likely part of the least-cost path (optimal traffic control plan) in the next Lagrangian decomposition iteration, resulting in those red light runners more likely meeting greens when they enter the intersections. Over iterations, the subgradients will gradually become small and ideally become zero eventually (i.e., reach the global optimum).

<u>Algorithm Two: Lagrangian relaxation and subgradient method</u>

In summary, the Lagrangian-relaxation-based optimization algorithm for $Z_1$ can be described as follows:

Step 1: Initialization

 1.1 Construct the space-time network according to the target road network

 1.2 Call Algorithm 1 to generate the set of generalized phases

 1.3 Construct the phase-time network according to the generalized phases and the corresponding traffic signal mandates (e.g., minimum green, maximum green or specified phasing sequence) 1.4 Construct the mapping matrix between the generalized phases and the corresponding traffic signal links in the target road network.

1.5 Choose the initial values for Lagrangian multipliers $\lambda_3(i,j,t,s), (i,j) \in A_s, t, s \sim [1, \mathcal{H}]$, such as 0

1.6 Read all vehicles' scheduled paths within the target road network

1.7 Set the maximal iterations of Lagrangian relaxations and the initial iteration number n = 0

Step 2: Call the customized DNL process, with neither traffic control effects nor capacity constraints on control links, to get the optimal solution to $\mathcal{L}_1^1$. Calculate the minimum of $\mathcal{L}_1^1$ according to the loading results and the latest $\lambda_3(i,j,t,s)$

Step 3: Calculate the phase-time arc costs with the latest values $\lambda_3(i,j,t,s)$ and (12') and then seek the shortest path in $\Psi(P,T)$ as the optimal solution to $\mathcal{L}_1^2$. The optimal solution is then translated into the latest network traffic control strategy for the target road networks.

Step 4: Calculate the lower bound (LB) of $Z_1$: $(\mathcal{L}_1^1 + \mathcal{L}_1^2)$

Step 5: Conduct the standard dynamic network loading again with the latest network traffic control strategy superposed to get the upper bound (UB) of $Z_1$

Step 6: Calculate the subgradient with (19), (20) and (21)

Step 7: $\lambda_3(i,j,t) = \lambda_3(i,j,t) + \max\left(0, \theta \times \nabla L_{\lambda_3(i,j,t)}\right), (\forall (i,j) \in A_s, t \sim [0, H])$; $\theta$ is the step size and equal to $\frac{1}{n+1}$

Step 8: If the maximal number of iterations is reached or the gap between LB and UB is smaller than the threshold, then stop and take the latest network traffic control strategy as the final solution. Otherwise go to Step 2.

## 7. Numerical Experiments

We conduct two experiments in this study to validate the proposed MILP formulation under various scenarios using a demonstrative example. For the first demonstrative experiment, all the results are visualized according to the solutions from the GAMS MIP solver (Bussieck and Meeraus, 2004).

### 7.1 Experiment One: Solving network traffic control problem using the GAMS MIP solver

The demonstrative example of Experiment One is composed of a two-intersection arterial network composed of 28 nodes, 14 regular links and 8 controlled links whose free-flow travel times are marked in Fig. 6. The link capacities are uniformly set to 1,800 vehicles per hour per lane and all links have two lanes. The link storage is set to 10 vehicles on long links and 2 vehicles on short links. Without loss of generality, turning movements are not configured and there are four network phases for this example. The initial configurations for all four phases are the same as shown in Fig. 6.

There are 20 vehicles in the network and 10 of them move along the mainline while the rest vehicles move on the side streets. $v_1$ to $v_{10}$ depart from node 1 and arrive at node 6, $v_{11}, v_{12}$ depart from node 17 and arrive at node 20, $v_{13}, v_{14}$ depart from node 13 and arrive at node 16, $v_{15}, v_{16}$ depart from node 25 and arrive at node 28, $v_{17}, v_{18}$ depart from node 21 and arrive at node 24, $v_{19}, v_{20}$ depart from node 7 and arrive at node 12. Through adjusting the departure times from all directions and link storage constraints, four representative traffic scenarios are created: (a) uncongested traffic with low competition between the main line and side streets; (b) uncongested traffic with high competition between the mainline and side streets; (c) congested traffic (queue spillback) with low competition between the mainline and side streets; and (d) congested condition (queue spillback) with high competition between the mainline and side streets. The minimum and maximum greens for generalized phases are derived from local phases at two intersections following Algorithm One.

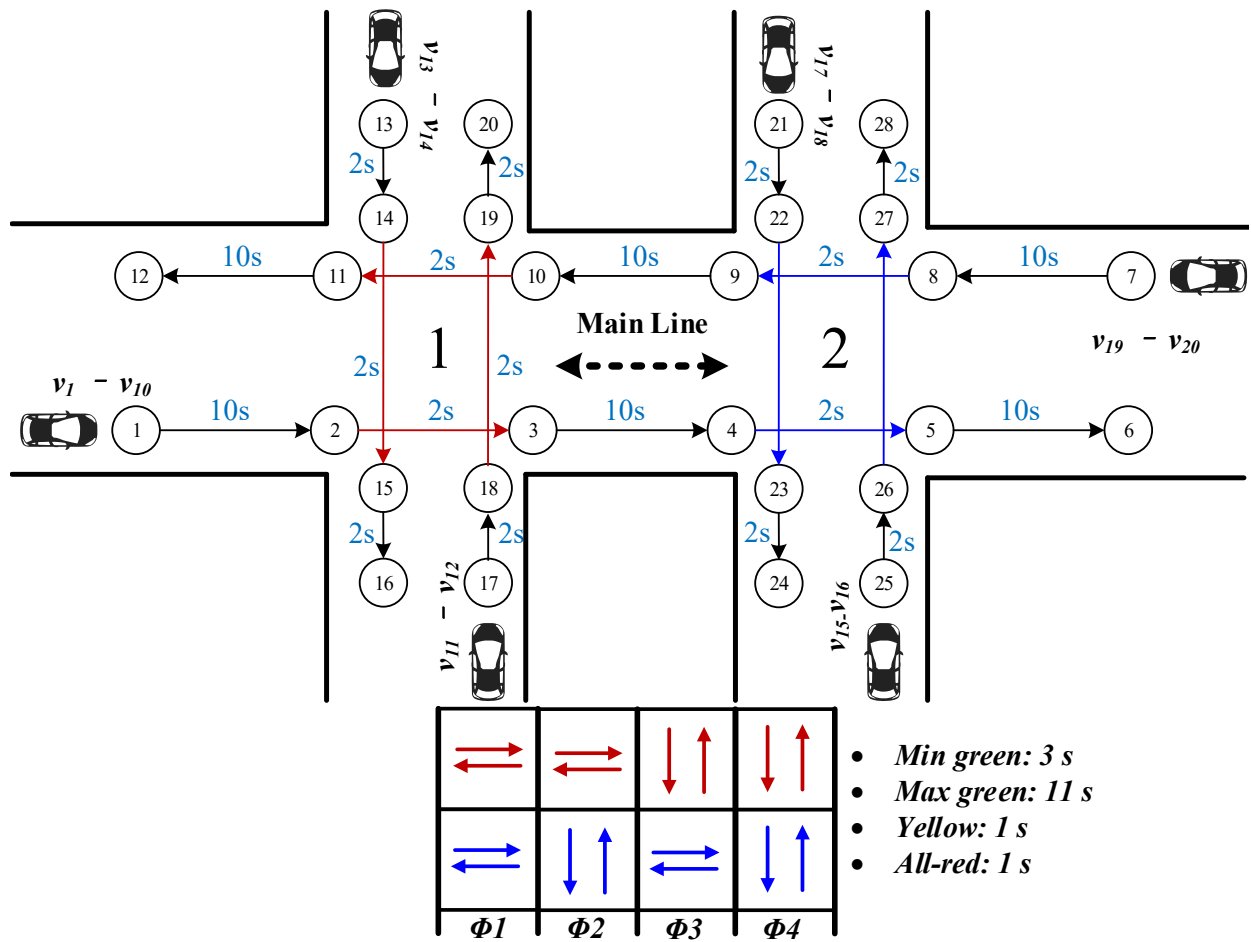

Figure 6 Network configuration for Experiment One

*Scenario One: Uncongested traffic and low competition between mainline and side streets*

In this scenario, the departure times of all vehicles are set in such a way that they will not compete for the green lights while crossing intersections. The traffic control solution and the resulting vehicle trajectories are revealed in Fig. 7. The GAMS MIP solver provides a solution allowing all vehicles to finish their paths at the free-flow speeds and it is the theoretical lower bound of the system optimum. Through the multithreading techniques (4 threads), the GAMS MIP solver takes 1.01 seconds to solve this problem and use 120MB memory to generate 9,284 rows, 176,070 columns, 656,336 non-zeroes and 176,069 discrete-columns while searching the optimal solutions.

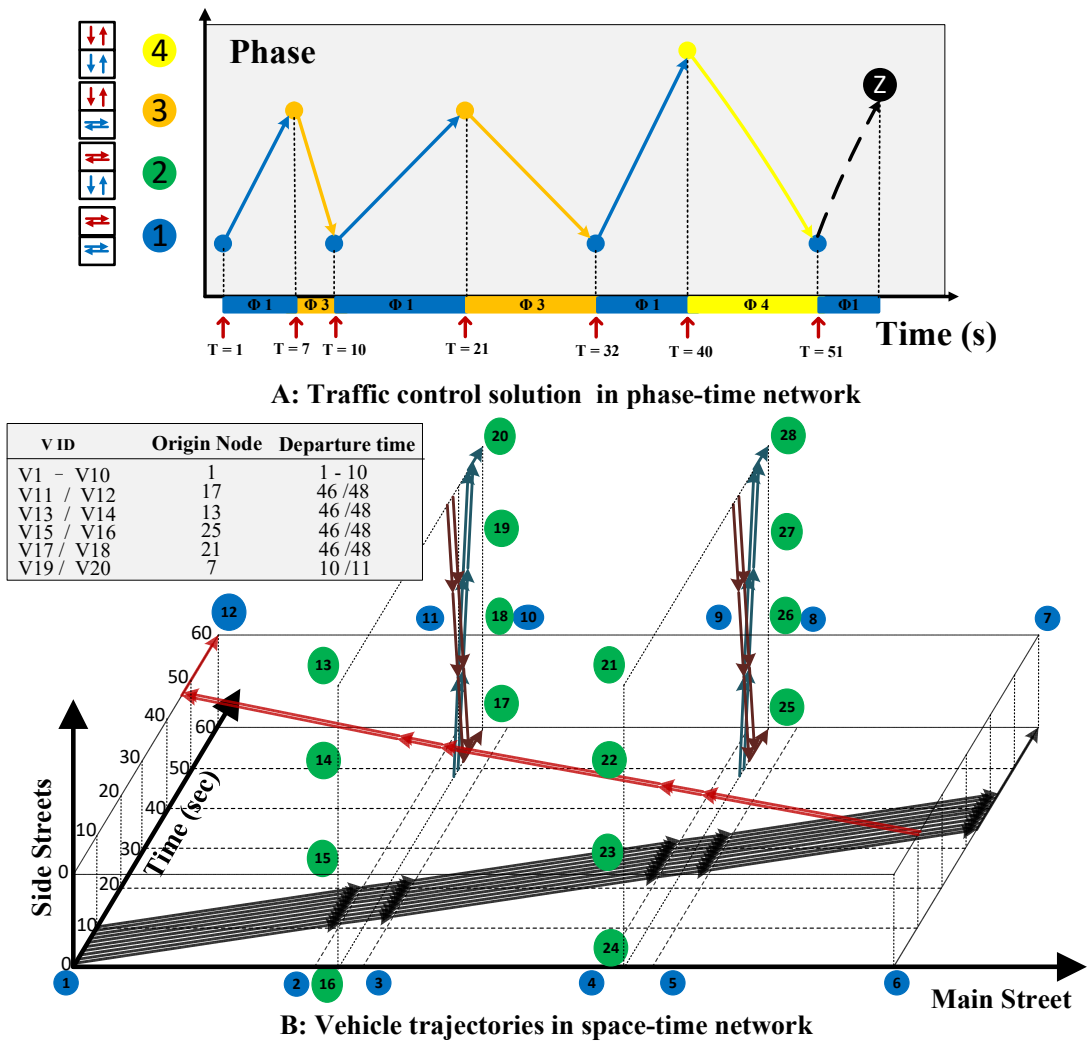

Figure 7 Optimal traffic control solution for uncongested traffic with low competition for green lights

*Scenario Two: uncongested traffic and high competition between mainline and side streets*

In this scenario, the departure times of all vehicles are set in such a way that they will compete for the green lights while crossing intersections. The traffic control solution and the resulting vehicle trajectories are revealed in Fig. 8. The GAMS MIP solver provides a solution allowing main street vehicles to finish their paths at the free-flow speeds and keep side street vehicle wait until all main street vehicles being cleared. The GAMS MIP solver takes 3.14 seconds to solve this problem and use 122MB memory to generate 9,284 rows, 176,070 columns, 656,336 non-zeroes and 176,069 discrete-columns while searching the optimal solutions.

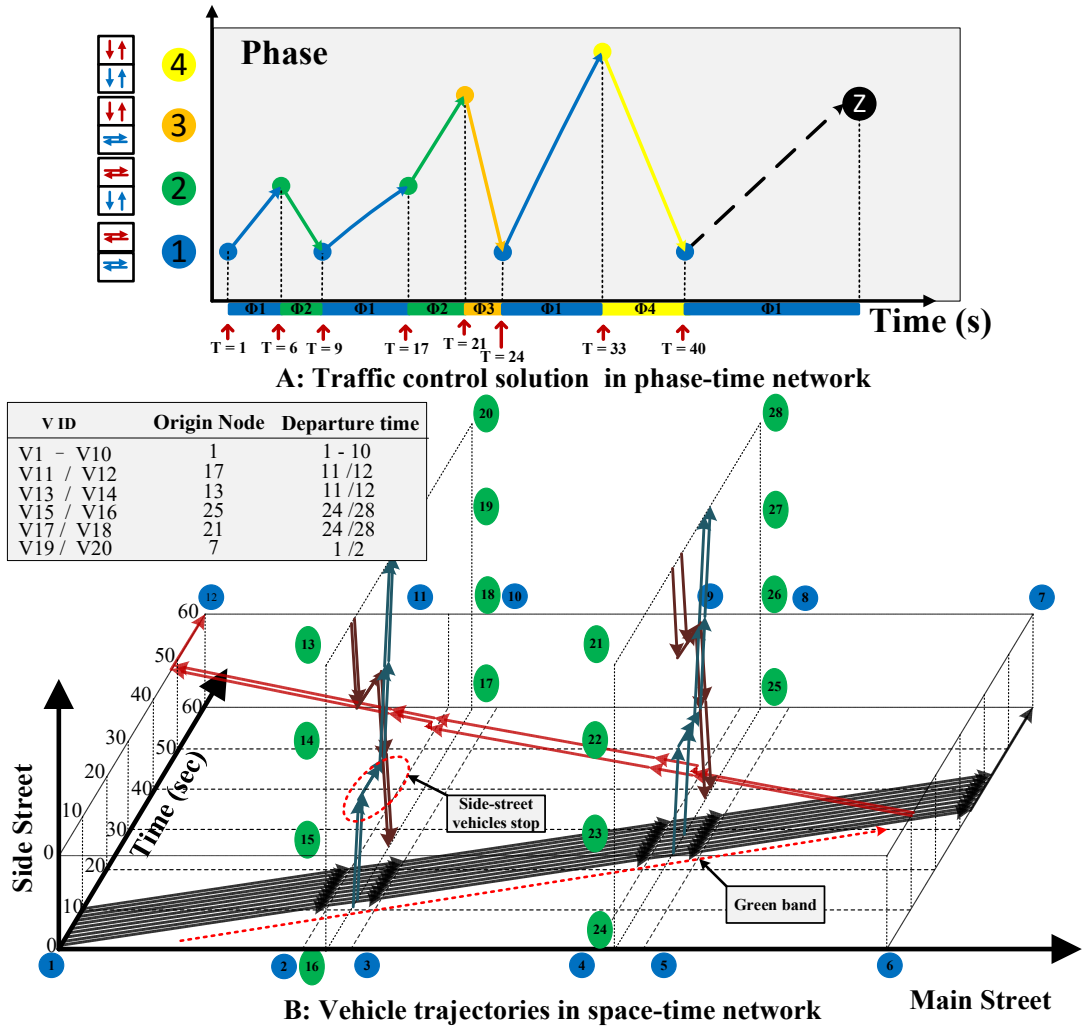

Figure 8 Optimal traffic control solution for congested traffic with high competition at intersections

*Scenario Three: Congested traffic (queue spillback) and low competition between mainline and side streets*

In this scenario, the departure times of all vehicles are set in such a way that they will compete for the green lights while crossing intersections. Compare to Scenario One, we reduce the link storage constraints on the main between two intersections ($9 \rightarrow 10, 3 \rightarrow 4$) from 10 vehicles to 5 vehicles. As a result, the solver has to split the EB green wave and hold the last five vehicles for a few seconds before they can cross the first intersection. The side street vehicles are not affected before they depart after the congestion is cleared. The traffic control solution and the resulting vehicle trajectories are revealed in Fig. 9. The solver takes 1.12 seconds to solve this problem and use 120MB memory to generate 9,284 rows, 176,070 columns, 656,336 non-zeroes and 176,069 discrete-columns while searching the optimal solutions.

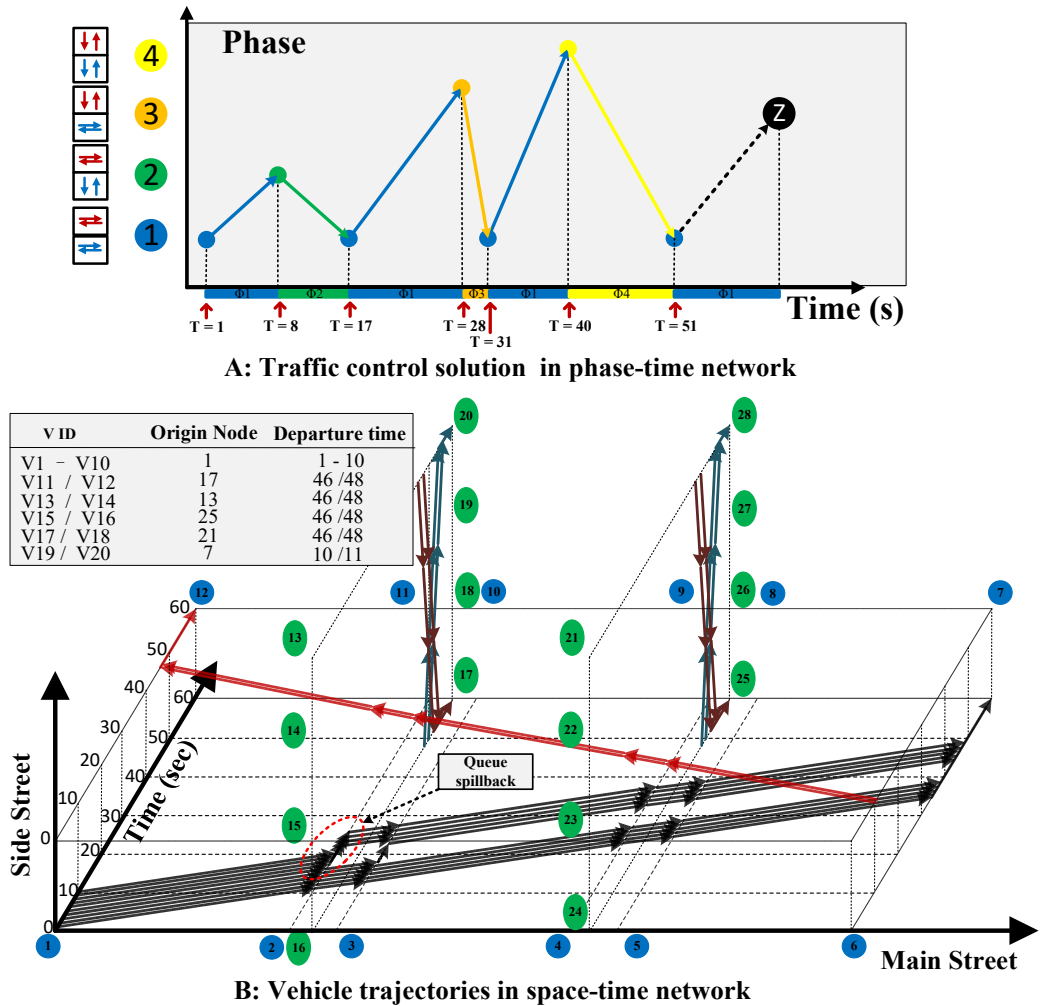

**A: Traffic control solution in phase-time network**

**B: Vehicle trajectories in space-time network**

Figure 9 Optimal traffic control solution for congested traffic with high competition for green lights

*Scenario Four: Congested traffic (queue spillback) and low competition between mainline and side streets*

Scenario Four represents the most challenging traffic conditions and it is in essence a combination of Scenario Two (high competition for crossing intersections) and Scenario Three (Mainline congestion). Fig. 10. Shows that the solver provides a solution allowing main street vehicles to finish their paths as fast as possible even some of them are held due to the queue spillback while the side street vehicle yield to the mainline traffic. the solver takes 3.22seconds to solve this problem and use 122MB memory to generate 9,284 rows, 176,070 columns, 656,336 non-zeroes and 176,069 discrete-columns while searching the optimal solutions.

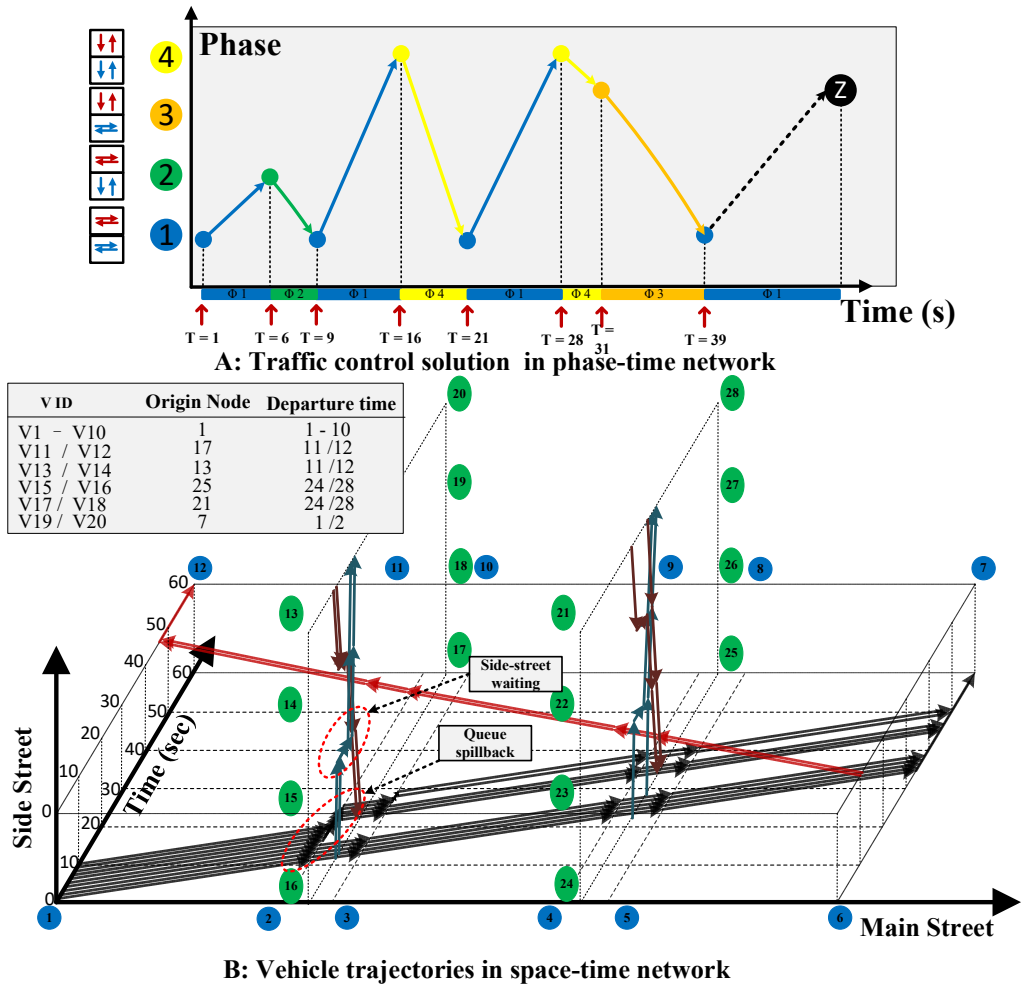

Figure 10 Optimal traffic control solution for congested traffic with high competition for green lights

*Discussion*

The four scenarios in Experiment One almost represent all common traffic control conditions. The GAMS MIP solver provides valid as well as reasonable solutions for all four scenarios and therefore we conclude that the proposed MILP formulation for network traffic control optimization is valid and robust. While the solver is seeking the solutions, it is noticed that the GAMS solver allows for using the multi-threading technique to improve the search efficiency while the memory usage under the different number of threads is nearly unchanged. We also notice that computing time is three times longer if vehicles must compete for the green lights to cross intersections than those low-competition scenarios while the computing efficiency is nearly unchanged if traffic is congested. Therefore, the computing complexity in the proposed MILP formulation likely lies on the side of the phase-time network.

We also compare the proposed MILP formulation with the well-known MAXBAND method under the same traffic condition and demonstrate that all the solutions from MAXBAND can be represented in the generalized phase time network and the proposed MILP formulation can provide better solutions than the MAXBAND in some circumstances because the proposed MILP formulation considers all vehicles while the MAXBAND only consider the mainline. The analysis is elaborated in the appendix A.

**8.2 Experiment Two: Traffic signal optimization along the Peachtree Street, Atlanta, GA, USA**

In the second experiment, we perform the traffic signal optimization for a real-world arterial, the Peachtree Street in Atlanta using the proposed scalable "Lagrangian relaxation plus subgradient" method. A complete set of traffic data collection, including traffic signal timing, turning movement counts and individual vehicle trajectories along this arterial, was conducted under the NGSIM program (Bared and Zhang, 2016) funded by the US Federal Highway Administration. The selected arterial contains four signalized intersections and there are heavy traffic on the side streets at Intersection 1 and 5. Therefore traditional traffic signal coordination, such as maximizing the green band along the mainline, is not efficient for this example. During 15 minutes of data collection, 1,127 vehicles' trajectories were captured entering the scope of arterial. To allow all vehicles to leave the network (i.e., to generate a feasible solution), the time horizon is set as 1,400s. The final best solutions are selected after 200 iterations.

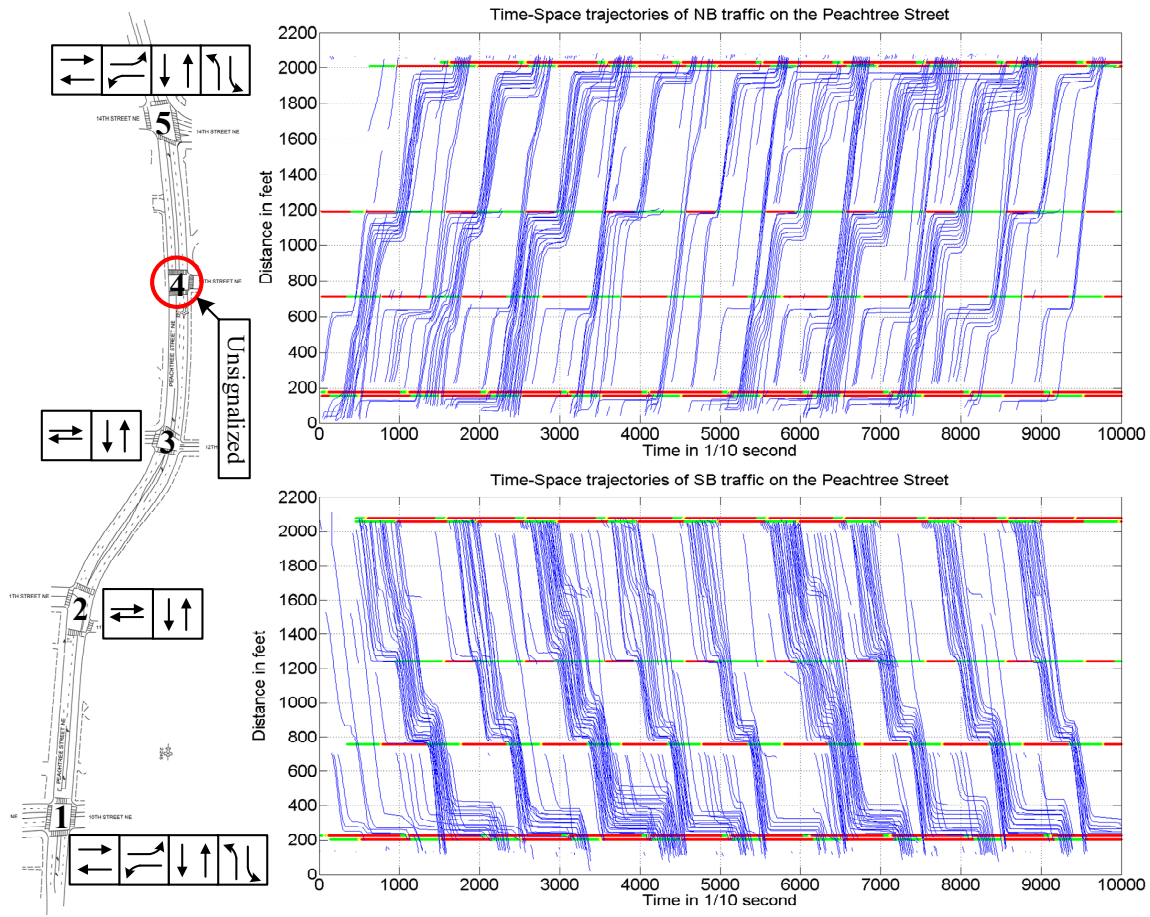

Figure 11 Example configurations for the real-world experiment

*Traffic control phase configurations*

Without loss of generality, phases at individual intersections are set as in Fig.1 and the minimum green, maximum green, yellow and all-red clearance for all local phases are set as: 5s, 50s, 4s and 3s respectively. As a result, 64 generalized phases are generated according to Algorithm One. Only one generalized phase can be activated at any time.

*Measures of effectiveness (MOE)*

To achieve the system optimum (minimal total delay), we adopt a novel MOE to evaluate the centralized network traffic control strategy after the optimization, namely "vehicle arrivals during greens". Since our MILP formulation is centered on vehicles and the final solution contains both traffic control plan and the corresponding all individual vehicles' space-time trajectories. The rational of "vehicle arrivals during greens" is that it is the most direct MOE for system optimum because the total control delays will be minimized if most vehicles can cross intersections during green without stops. Compared with the total travel time, the arrivals during green also reveals more in-process information.

*Scenario One: Completely adaptive network traffic control strategy*

In this scenario, the optimization algorithm is allowed to switch from one phase to any other phase if it can find a better solution to reduce the total travel delays. This scenario represents a completely adaptive network traffic control strategy to reach the lower bound of system optimum. As a result, there are totally 4,032 feasible phase transitions. This scenario is further divided into two sub-scenarios: with permissive turnings and without permissive turnings. Although the permissive vehicle turnings are ubiquitous in some regions like North America, most traditional traffic control models only address the capacity issue of protected turning movements. The permissive turning phases are explicitly modeled in the proposed MILP formulation and scalable optimization algorithm. From Fig. 12 and Fig. 13, we can tell that most vehicles on the main line arrive at intersections during the green while the solution algorithm holds individual vehicles from side streets for very long times to minimize the total delay. In reality, excessive waiting may lose drivers' respect to the traffic control system, and therefore this solution is not realistic even though it can achieve good system optimum.

In the subject network, all intersections allow for permissive left turns. Between to sub-scenarios, the total travel time with permissive left turn allowed is 51,012s as opposed to the 80,450s without permissive left turns, interpreted as a 57.7% overestimation of total travel time if the intersection capacities during the permissive left turn are ignored in developing the network traffic control strategy.

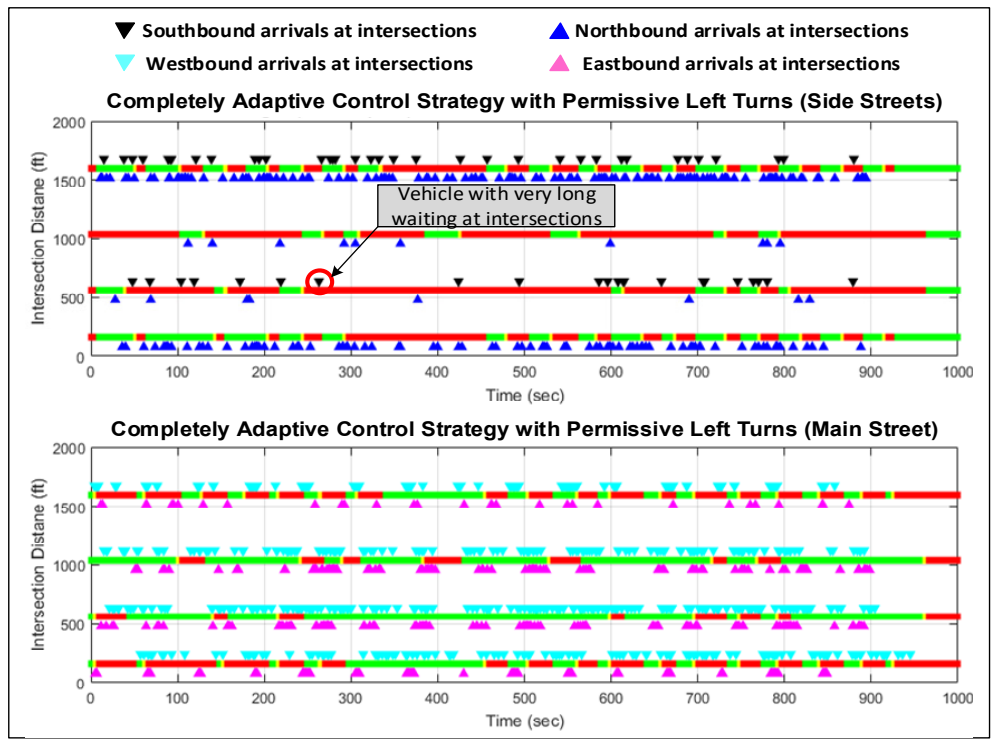

Figure 12 Fully adaptive control solution with permissive left turns and vehicle arrivals at intersections

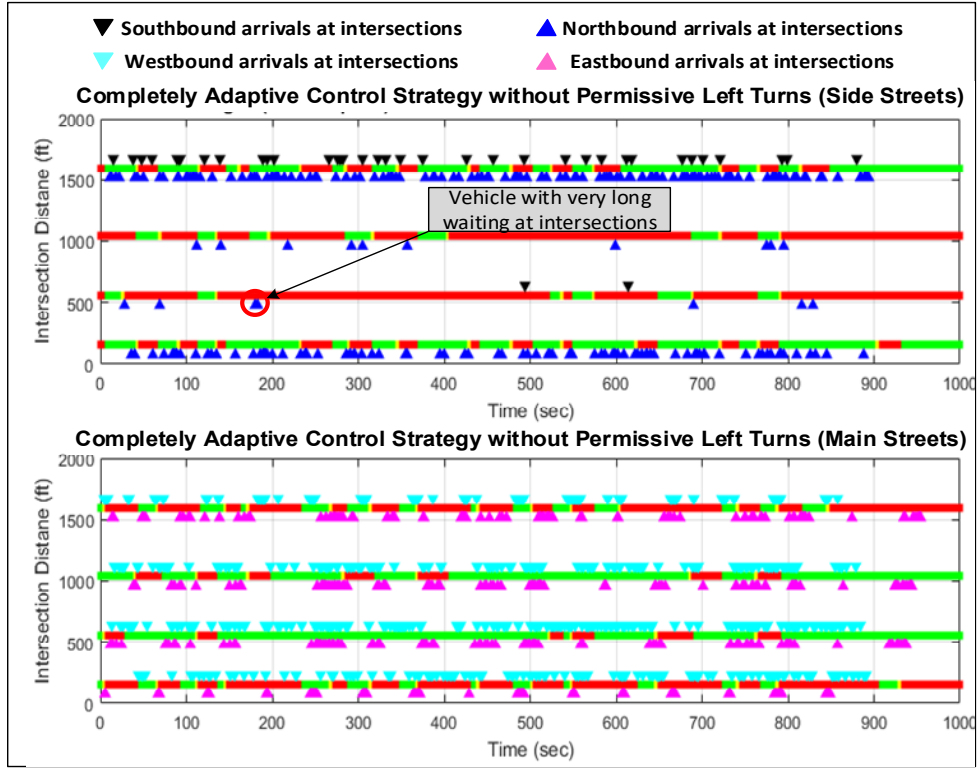

Figure 13 Fully adaptive control solution without permissive turns and vehicle arrivals at intersections

*Scenario Two: Semi-adaptive network traffic control strategy*

In some regions, the local phasing sequence at individual intersections is not allowed to change. In this scenario, we modify the phase transition rule to reflect this requirement. Let a string "$P_1P_2P_{3,\dots}P_i \dots P_n$" denote a generalized phase which is composed of one and only one local phase $P_i$ at intersection $i$, ($i = 1, 2, \dots, n$). If a current generalized phase is "$P_1^1P_2^1P_{3,\dots}^1P_i^1 \dots P_n^1$" and the phasing sequence requirement at intersection $i$ is fixed as $P_i^1 \rightarrow P_i^2$, then the feasible next generalized can only be "$Q_1Q_2Q_{3,\dots}Q_i \dots Q_n$" where $Q_i = P_i^1$ (local phase is unchanged) or $P_i^2$ (local phase switch to the next phase). Following this rule, the total number of phase transitions is reduced to 1,984 after we set local phase sequences as ($\Phi1 \rightarrow \Phi2 \rightarrow \Phi3 \rightarrow \Phi4 \rightarrow \Phi1 \dots$) at intersection 1 and 5 or ($\Phi1 \rightarrow \Phi2 \rightarrow \Phi1 \dots$) at intersection 2 and 3. From Fig. 14 and Fig. 15, we can see that the final solution is also effective in maintaining the traffic mobility on the main line while the issue of excessive waiting on the side streets still exists because the transition rule allows the local phase unchanged between two generalized phases. As a result, this control strategy cannot effectively prevent the long-waiting issues on the side streets.

In the meantime, the total travel times with and without a permissive left turn in this scenario slightly increase compared with the completely adaptive control strategy: 56,305s as opposed to 51,012s (with permissive left turns) and 86,704s as opposed to 80,450s (without permissive left turns). Considering the search space is significantly reduced from a completely adaptive control strategy to the semi adaptive control strategy (51% in this example), the semi-adaptive control strategy should be preferred over the completely adaptive control strategy in most cases.

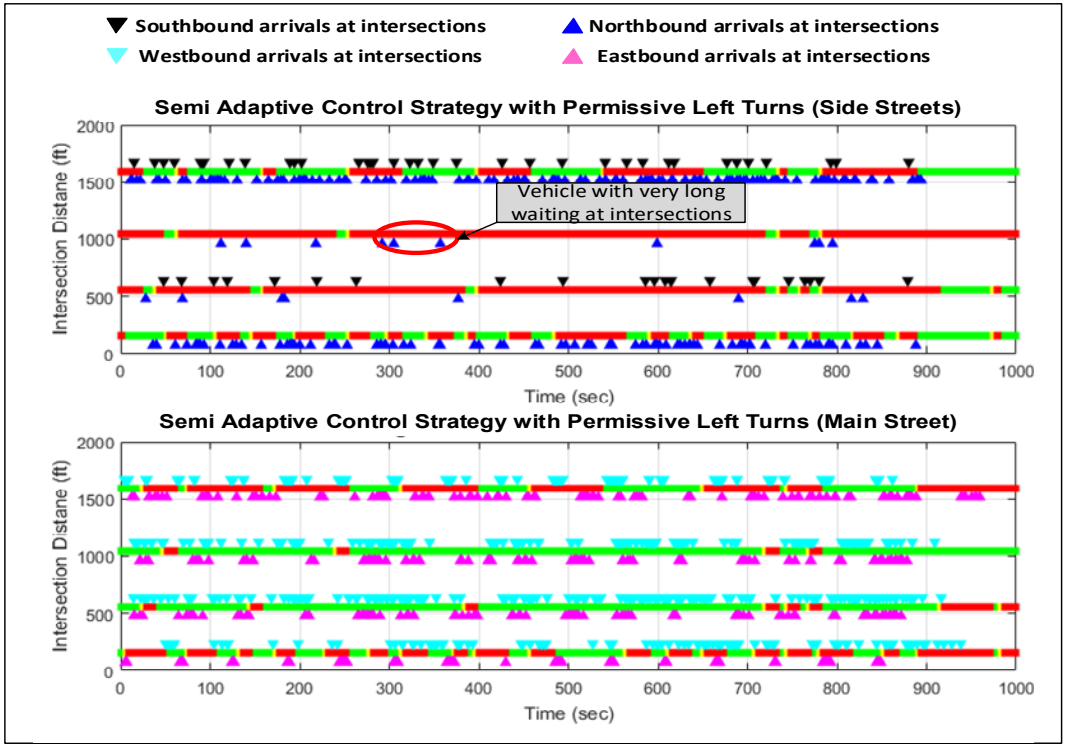

Figure 14 Semi adaptive control solution with permissive left turns

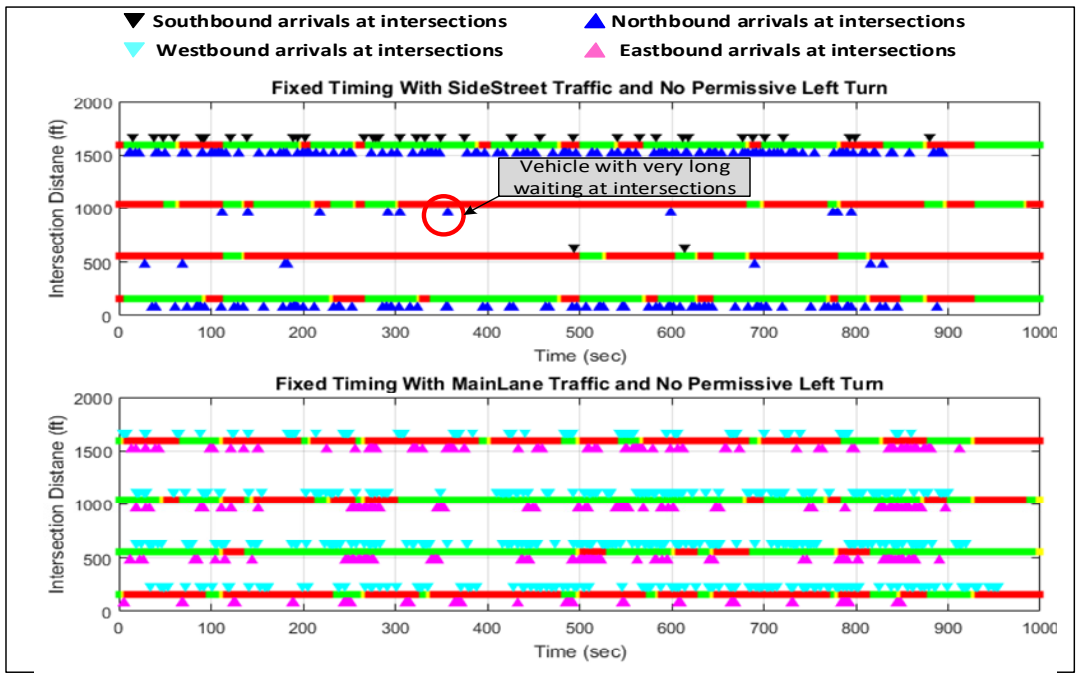

Figure 15 Semi adaptive control solution without permissive left turns

*Scenario Three: semi adaptive control strategy with local maximum green restrictions*

In this scenario, we superimpose maximum greens for each local phase. The maximum greens of local phases are not modeled in the proposed MILP formulation. Instead, such restrictions are superimposed the label-based search algorithms for the shortest path in the generalized phase network which is elaborated in Appendix B.

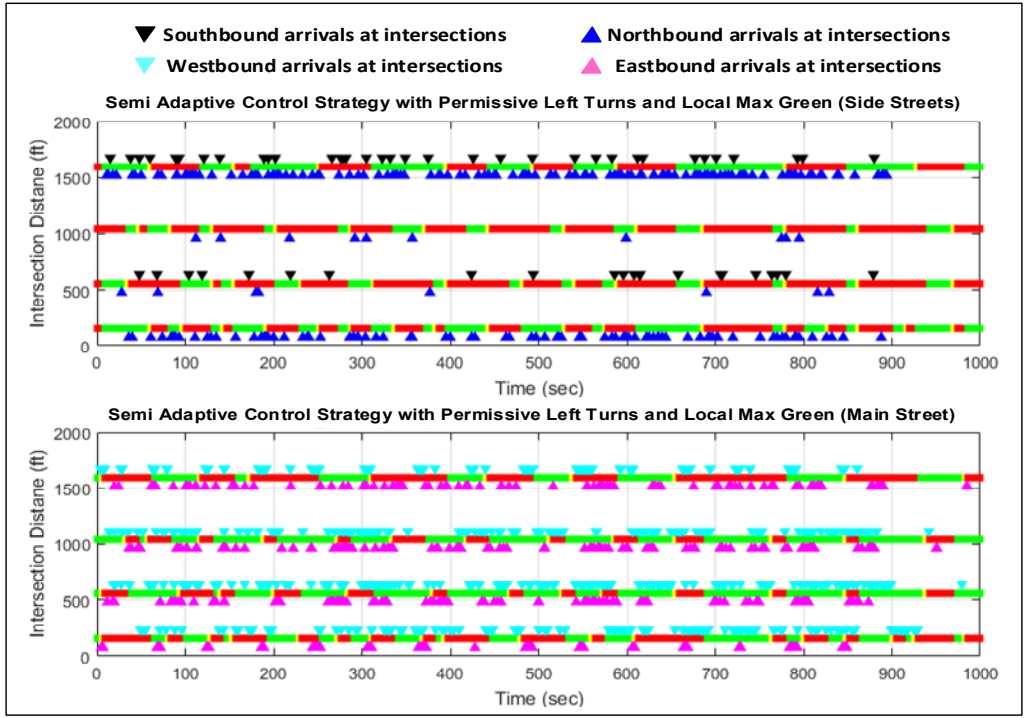

Figure 16 Semi adaptive control solution with permissive left turns and local max greens

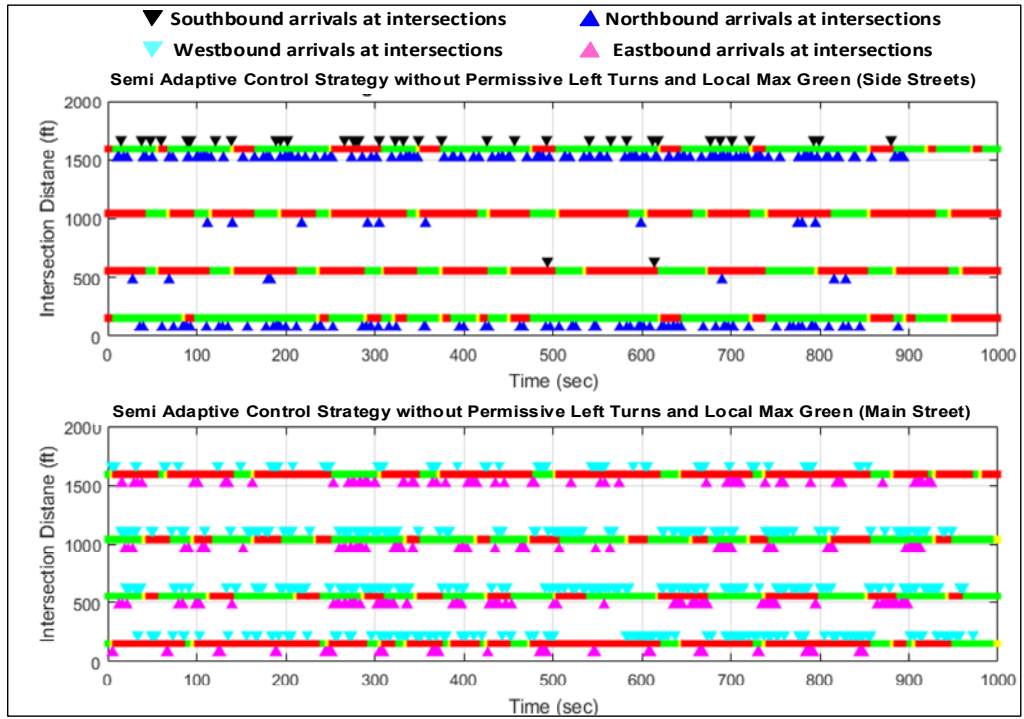

Figure 17 Semi adaptive control solution without permissive left turns and local max greens

*Performance evaluation of HPC technique*

The multi-threading technique is deployed during the optimization. In each iterations, the following computing tasks can be parallelized:

1. Special DNL in which the control links' capacity constraints are dropped. (part of lower bound)

2. Shortest path finding for the optimal traffic control strategy in the phase-time network (part of lower bound)

3. Updating the Lagrangian multipliers (bundles between two subproblems in the relaxed problems)

4. Standard DNL under the optimal traffic control strategy to get a feasible solution (upper bound)

Fig. 18 reveals the average computing time for each iteration with various threads (i.e., the number of dedicated CPU cores). We can clearly see the significant computing time reduction from a few CPU cores to more while the marginal performance improvement decreases after the number of CPU cores are more than 20. This is expected because the synchronization between threads comes to govern the computing speed while the number of threads keeps increasing. It is also noticed that the computing time per iteration increase if the permissive turning is allowed due to the increase of search efforts for the permissive turning movements.

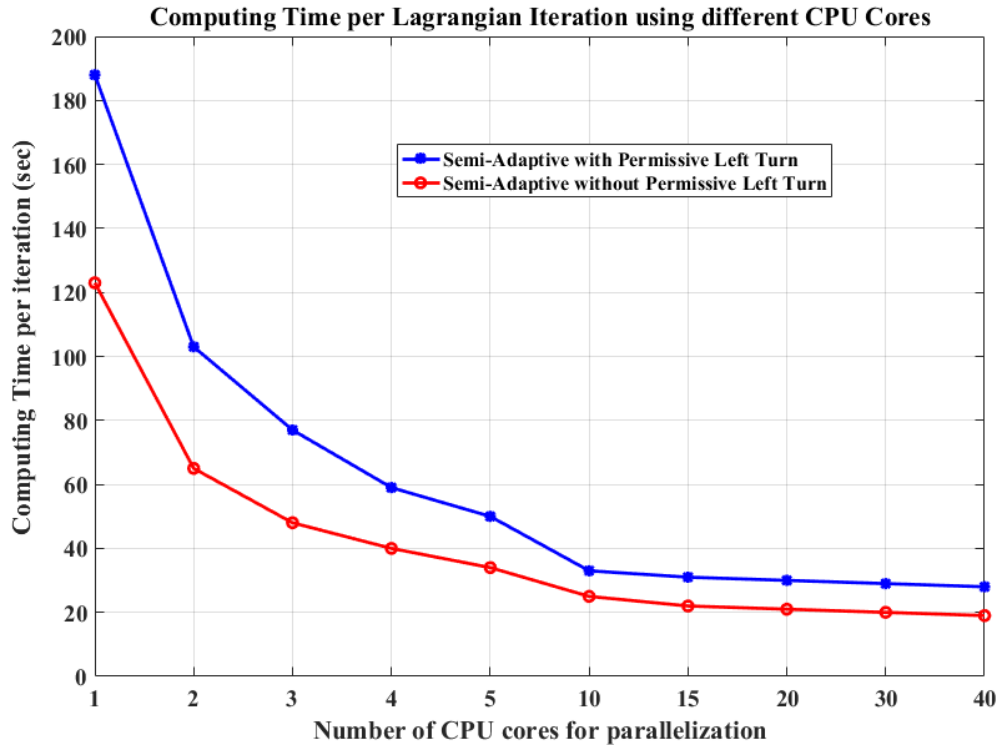

Figure 18 Computing performance evaluation under different CPU cores

## 9. Conclusion and future work

The crux of this paper is to provide a new centralized adaptive traffic control representation based on vehicle path flows at the age of mobile computing. In foreseeable future, vehicle path flows will become ubiquitous and provide full-spectrum space-time connections among multiple intersections. Vehicles are sharing their real-time states today and it will allow us to develop highly centralized traffic control strategies with enhanced flexibility and robustness. Inspired by these promising trends, the proposed centralized traffic control strategy is based on the path flows across multiple intersections and allow all intersections to closely coordinate in order to minimize the total travel times. The proposed centralized traffic control strategy can also be used to maximize the green bands efficiently if we fix the sequence and durations of a group of generalized phases in optimization as illustrated in Fig.2.

We also propose a new approximated Lagrangian decomposition solution. Unlike the classic mathematical formulations for optimization problems, we reformulate the target problem by constructing most of the target problem's complicating attributes into the objective function. As a result, the objective function of the transformed problem objective function becomes of a dynamic network loading (DNL) type while the computing complexity is significantly reduced. Multiple high-performance computing techniques can be applied to mitigating the loss of computing efficiency on the side of objective function. The transformed problem can be solved with the proposed approximated Lagrangian decompose solution based on one standard DNL model and one customized DNL model. We consider this approximated Lagrangian decomposition solution can be generalized for other large-scale time-dependent transportation problems.

Through a demonstrative example, the proposed MILP formulation can generate solutions with the GAMS MIP solver efficiently and reasonably in various scenarios representing most traffic control conditions. We also solve a real-world problem based on the above approximated Lagrangian Decomposition method.

In the future, we plan to investigate additional optimization approaches other the subgradient method for the proposed optimization framework to further improve the search efficiency. In the meanwhile, we plan to investigate approximate dynamic programming (DP) approaches in searching the shortest path in the phase-time network as approximate search will become necessary when the network scope becomes large.

## 10. Acknowledgement

This study is partially supported by the project titled "*Developing HPC-enabled Simulation System for Multi-scale Mobility Network Analytics*" funded by the Institute of Systems Engineering Research (ISER) at Mississippi State University. The authors would also thank Messrs. Randy Jones and T.C. Falls at Mississippi State University for offering the access to the on-campus HPC Collaboratory facilities. Any opinions, findings, and conclusions or recommendations expressed in this material are those of the authors and do not necessarily reflect the official views or policies of the above organizations, nor do the contents constitute a standard, specification, or regulation of these organizations.

## Appendix A: Comparison between MAXBAND and the proposed MILP formulation to optimize the network traffic control strategy

As required by the MAXBAND (Little et al., 1981), the cycle and splits at each individual intersections must be provided as inputs. Based on the traffic conditions of Scenario Two in Experiment One, the traffic volumes on the mainline and side streets can be interpreted as 1,200 vehicles per hour per lane and 120 vehicles per hour per lane respectively. The cycle length, split for the mainline and split for the side streets at two intersections are 48 s, 42 s, 6 s, which are the required inputs for the MAXBAND. The travel time between to intersections are 12 s as in Experiment One. The MAXBAND MILP formulation is also modelled in GAMS and the results are show in Fig. A1. $b, \bar{b}$ are the inbound and outbound bandwidths defined in MAXBAND. As shown in Fig. A1, the MAXBAND solution can be completely represented by a sequence of generalized phases.

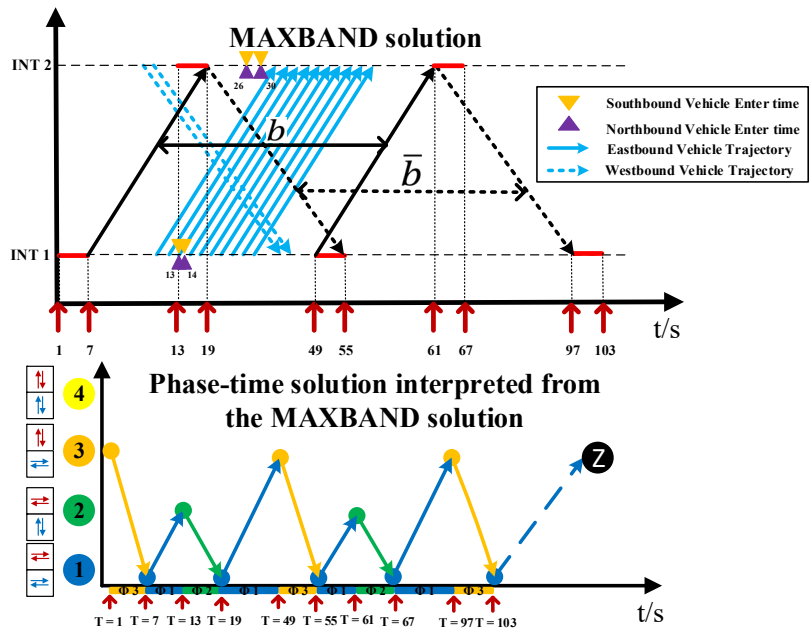

Figure A1, MAXBAND solution and its interpretation with the generalized phase-time networks

To be comparable with the MAXBAND solution, we fix the phasing sequence as $\Phi 1 = 30$ s or 6 s (mainline phase) and $\Phi 2 = \Phi 3 = \Phi 4 = 6$ s (with at least one side-street local phase). Fig. A2 reveals the GAMS solution based these settings. Since the proposed MILP formulation no longer requires a fixed cycle length, the solution is flexible to maximize the number of vehicles which cross two intersections without stops.

It should be pointed out that the phase-time solution out of the proposed MILP formulation is better than the corresponding MAXBAND solution even though the mainline solution by the MAXBAND is slightly better than the one by the proposed approach. The MAXBAND solution does not consider the 8 vehicles on the side streets and those 8 vehicles incur totally 274 (34 s per vehicle) seconds delays while the new phase-time solution considers all vehicles and therefore the 8 vehicles on the side streets can also cross the intersection with 32 sec delay and total 70 sec delay when consider all directions.

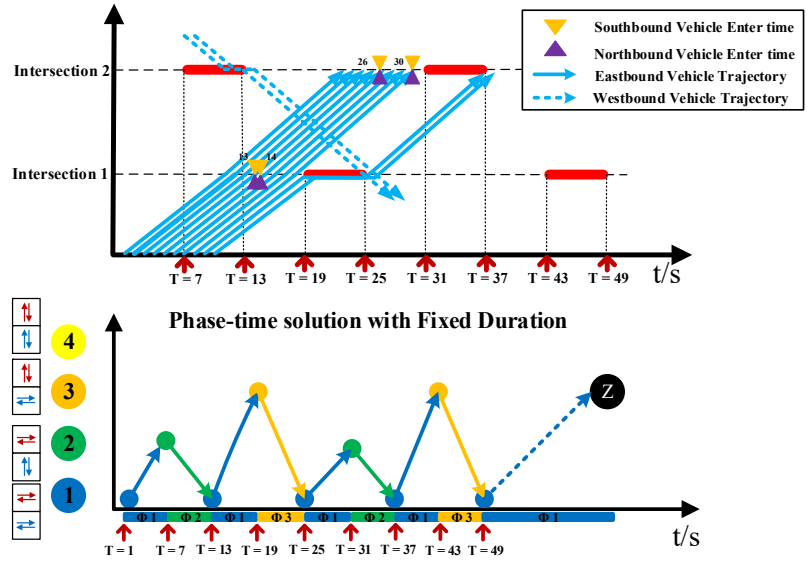

Figure A2, Solution out of the proposed MILP formulation and the resulting vehicle trajectories

## Appendix B: A labeling algorithm to optimize the centralized traffic control strategy considering the maximum greens of local phases.

Let $\Psi(P,T)$ denote a generalized phase time network; $\forall p, p' \in P$; $p_0$ denote the initial phase; $l_i^p$ denote a local phase which is part of the generalized phase $p$ at intersection $i$; $\tau, h$ denote time indices; $c(p, \tau, p', h)$ denote the phase-time arc cots; $lc(p, \tau)$ denote the label cost of $(p, \tau)$; $\max(p)$ denote maximum greens of generalized phase $p$; $\max(l_i^p)$ denote the maximum green of $p'$ corresponding local phase at intersection $i$; $\mathcal{F}(p, \tau) = \{l_1^p, l_2^p, \dots, l_i^p\}$ denote a collection of active local phases at each intersection $(1, 2, \dots, i)$ right before $p$ is activated; $\mathcal{R}(p, \tau) = \{t(l_1^p), t(l_2^p), \dots, t(l_i^p)\}$ denote the collection of total durations of the active local phases at each intersection $(1, 2, \dots, i)$ right before $p$ is activated; $T$ denote the time horizon and; $Pred(p, \tau)$ denote the predecessor of $(p, \tau)$ in the shortest path (optimal traffic control operations) in $\Psi(P, T)$. The polynomial labeling algorithm can be described as follows.

Initialization: $lc(p, t) = +\infty, (t = 0, 1, \dots, T)$; $lc(p_0, 0) = 0$; $\mathcal{F}(p_0, 0) = \{l_1^{p_0}, l_2^{p_0}, \dots, l_i^{p_0}\}$; $\mathcal{R}(p, \tau) = \{t(l_1^{p_0}) = 0, t(l_2^{p_0}) = 0, \dots, t(l_i^{p_0}) = 0\}$

FOR $\tau = 0$ to $T\text{-}1$

  FOR $\forall p \in P$

   FOR $\forall p' \in P$ (IF $p \rightarrow p'$ is a feasible transition)

   FOR $\Delta t = \min(p)$ to $\max(p)$

   IF $(lc(p', \tau + \Delta t) > lc(p, \tau) + c(p, \tau, p', \tau + \Delta t)$

    FOR all the elements $\mathcal{F}(p, \tau) = \{l_1^p, l_2^p, \dots, l_i^p\}$

     IF $\left(l_i^p = l_i^{p'}\right)$ // the local phase at intersection $i$ will stay green when $p \rightarrow p'$

      $t\left(l_i^{p'}\right) = t(l_i^p) + \Delta t$

     ELSE // a new local phase at intersection $i$ will be activated when $p \rightarrow p'$

      $t\left(l_i^{p'}\right) = 0$

     END //finish checking the local maximum green requirement;

    END //finish scanning all the local phases

    IF $\forall\, t\left(l_i^{p'}\right) \in \mathcal{F}(p, \tau) \leq \max\left(l_i^{p'}\right)$ for all intersections

     $lc(p', \tau + \Delta t) = lc(p, \tau) + c(p, \tau, p', \tau + \Delta t)$

     $Pred(p', \tau + \Delta t) = (p, \tau)$

    END //finish updating the label cost and predecessors

   END //finish checking each intersections

   END // for each $\Delta t$

  END //for each $p'$

 END //for each $p$

END //for each $\tau$